\documentclass[sn-basic]{sn-jnl}

\usepackage{graphicx}%
\usepackage{multirow}%
\usepackage{amsmath,amssymb,amsfonts}%
\usepackage{amsthm}%
\usepackage{mathrsfs}%
\usepackage[title]{appendix}%
\usepackage{xcolor}%
\usepackage{textcomp}%
\usepackage{manyfoot}%
\usepackage{booktabs}%
\usepackage{algorithm}%
\usepackage{algorithmicx}%
\usepackage{algpseudocode}%
\usepackage{listings}%
\usepackage{tikz}
 
\usepackage{amssymb}
\usepackage{pifont}
\newcommand{\cmark}{\ding{51}}%
\newcommand{\xmark}{\ding{55}}%

\usepackage[normalem]{ulem}

\newcommand{\msun}{\mbox{$M_{\odot}$}}

\newcommand{\rpro}{\mbox{{\it r}-process}}

\begin{document}

\title[ANDES Stars Science Case]{The discovery space of ELT-ANDES. Stars and stellar populations}

\author*[IUR1,IUR2]{\fnm{Ian U.} \sur{Roederer}}\email{iur@umich.edu}
\author[JAG]{\fnm{Juli\'an D.} \sur{Alvarado-G\'omez}}\email{julian.alvarado-gomez@aip.de}
\author[CAP1,CAP2]{\fnm{Carlos} \sur{Allende Prieto}}\email{callende@iac.es}
\author[VA]{\fnm{Vardan} \sur{Adibekyan}}\email{vadibekyan@astro.up.pt}
\author[DA1,DA2]{\fnm{David} \sur{Aguado}}\email{david.aguado@iac.es}
\author[PJA]{\fnm{Pedro J.} \sur{Amado}}\email{pja@iaa.csic.es}
\author[EMAG]{\fnm{Eliana M.} \sur{Amazo-G\'omez}}\email{eamazogomez@aip.de}
\author[MBA1,MBA2]{\fnm{Martina} \sur{Baratella}}\email{Martina.Baratella@eso.org}
\author[SAB]{\fnm{Sydney A.} \sur{Barnes}}\email{sbarnes@aip.de}
\author[TB]{\fnm{Thomas} \sur{Bensby}}\email{thomas.bensby@fysik.lu.se}
\author[LB]{\fnm{Lionel} \sur{Bigot}}\email{lionel.bigot@oca.eu}
\author[AC]{\fnm{Andrea} \sur{Chiavassa}}\email{andrea.chiavassa@oca.eu}
\author[AD]{\fnm{Armando} \sur{Domiciano de Souza}}\email{armando.domiciano@oca.eu}
\author[CJH]{\fnm{Camilla Juul}
\sur{Hansen}}\email{hansen@iap.uni-frankfurt.de}
\author[SPJ]{\fnm{Silva P.} \sur{J\"arvinen}}\email{sjarvinen@aip.de}
\author[AJK]{\fnm{Andreas J.}\sur{Korn}\email{andreas.korn@physics.uu.se}}
\author[SL]{\fnm{Sara}\sur{Lucatello}\email{sara.lucatello@inaf.it}}
\author[LM]{\fnm{Laura} \sur{Magrini}}\email{laura.magrini@inaf.it}
\author[RM]{\fnm{Roberto} \sur{Maiolino}}\email{rm665@cam.ac.uk}
\author[PM]{\fnm{Paolo} \sur{Di Marcantonio}}\email{paolo.dimarcantonio@inaf.it}
\author[AMa]{\fnm{Alessandro} \sur{Marconi}}\email{alessandro.marconi@inaf.it}
\author[JRM]{\fnm{Jos\'e R.} \sur{De Medeiros}}\email{renan@fisica.ufrn.br}
\author[AM1,AM2]{\fnm{Alessio}\sur{Mucciarelli}\email{alessio.mucciarelli2@unibo.it}}
\author[NNa]{\fnm{Nicolas} \sur{Nardetto}}\email{Nicolas.Nardetto@oca.eu}
\author[LO]{\fnm{Livia}\sur{Origlia}\email{livia.origlia@inaf.it}}
\author[CP]{\fnm{Celine}\sur{Peroux}}
\email{celine.peroux@gmail.com}
\author[KP1, KP2]{\fnm{Katja} \sur{Poppenh\"ager}}\email{kpoppenhaeger@aip.de}
\author[CRL]{\fnm{Cristina} \sur{Rodr\'iguez-L\'opez}}\email{crl@iaa.es}
\author[DR]{\fnm{Donatella} \sur{Romano}\email{donatella.romano@inaf.it}}
\author[SN]{\fnm{Stefania} \sur{Salvadori}\email{stefania.salvadori@unifi.it}}
\author[PT]{\fnm{Patrick} \sur{Tisserand}}\email{tisserand@iap.fr}
\author[KV1]{\fnm{Kim} \sur{Venn}}\email{kvenn@uvic.ca}
\author[GAW]{\fnm{Gregg} \sur{Wade}}\email{wade.gregg@queensu.ca}
\author[AZ]{\fnm{Alessio} \sur{Zanutta}}\email{alessio.zanutta@inaf.it}


\affil[IUR1]{%
\orgdiv{Department of Astronomy}, 
\orgname{University of Michigan}, 
\orgaddress{%
\city{Ann Arbor}, 
\state{MI}
\postcode{48104}, 
\country{USA}}}

\affil[IUR2]{%
\orgdiv{Department of Physics}, 
\orgname{North Carolina State University}, 
\orgaddress{%
\city{Raleigh}, 
\state{NC}
\postcode{27695}, 
\country{USA}}}

\affil[JAG]{%
\orgdiv{Leibniz Institute for Astrophysics Potsdam (AIP)}, 
\orgaddress{
\city{Potsdam}, 
\postcode{14482}, 
\country{Germany}}}

\affil[CAP1, DA1]{%
\orgdiv{Instituto de Astrof\'{i}sica de Canarias}
\orgaddress{
\city{La Laguna},
\state{Tenerife}
\postcode{38205},
\country{Spain}}}

\affil[CAP2, DA2]{%
\orgdiv{Departamento de Astrof\'{i}sica}
\orgname{Universidad de La Laguna}
\orgaddress{
\city{La Laguna},
\state{Tenerife}
\postcode{38206},
\country{Spain}}}

\affil[VA]{%
\orgdiv{Instituto de Astrof\'isica e Ci\^encias do Espa\c{c}o, Universidade do Porto, CAUP, Rua das Estrelas, 4150-762}
\city{Porto}, 
\country{Portugal}}

\affil[PJA, CRL]{%
\orgdiv{Instituto de Astrof\'isica de Andaluc\'ia (IAA-CSIC)}, 
\orgaddress{%
\city{Granada}, 
\postcode{18008}, 
\country{Spain}}}

\affil[EMAG]{%
\orgdiv{Leibniz Institute for Astrophysics Potsdam (AIP)}, 
\orgaddress{%
\city{Potsdam}, 
\postcode{14482}, 
\country{Germany}}}

\affil[MBA1]{%
\orgdiv{Leibniz Institute for Astrophysics Potsdam (AIP)}, 
\orgaddress{
\city{Potsdam}, 
\postcode{14482}, 
\country{Germany}}}

\affil[MBA2]{%
\orgdiv{European Southern Observatory (ESO), Chile}, 
\orgaddress{
\city{Santiago}, 
\postcode{19001}, 
\country{Chiles}}}

\affil[SAB]{%
\orgdiv{Leibniz Institute for Astrophysics Potsdam (AIP)}, 
\orgaddress{%
\city{Potsdam}, 
\postcode{14482}, 
\country{Germany}}}

\affil[TB]{%
\orgdiv{Lund Observatory, Division of Astrophysics, Department of Physics}, 
\orgname{Lund University}, 
\orgaddress{%
\postcode{Box 43, 22100}, 
\city{Lund}, 
\country{Sweden}}}

\affil[LB]{%
\orgdiv{Universit\'e C\^ote d'Azur, Observatoire de la C\^ote d'Azur, CNRS, Lagrange}, 
\city{Nice}, 
\country{France}}

\affil[AC]{%
\orgdiv{Universit\'e C\^ote d'Azur, Observatoire de la C\^ote d'Azur, CNRS, Lagrange}, 
\orgaddress{%
\city{Nice}, 
\country{France}}}

\affil[AD]{%
\orgdiv{Universit\'e C\^ote d'Azur, Observatoire de la C\^ote d'Azur, CNRS, Lagrange}, 
\orgaddress{%
\city{Nice}, 
\country{France}}}

\affil[CJH]{%
\orgdiv{Goethe University Frankfurt, Institute for Applied Physics, 
\country{Germany}}}

\affil[SPJ]{%
\orgdiv{Leibniz Institute for Astrophysics Potsdam (AIP)}, 
\orgaddress{%
\city{Potsdam}, 
\postcode{14482}, 
\country{Germany}}}

\affil[AJK]{%
\orgdiv{Department of Physics and Astronomy}, 
\orgname{Uppsala University}, 
\orgaddress{%
\city{Uppsala}, 
\postcode{75120}, 
\country{Sweden}}}

\affil[SL]{%
\orgdiv{Osservatorio Astronomico di Padova}, 
\orgname{INAF}, 
\orgaddress{%
\city{Padova}, 
\postcode{35122}, 
\country{Italy}}}

\affil[LM]{%
\orgdiv{Osservatorio Astrofisico di Arcetri}, 
\orgname{INAF}, 
\orgaddress{%
\city{Firenze}, 
\postcode{I50125}, 
\country{Italy}}}

\affil[RM]{%
\orgdiv{Kavli Institute for Cosmology, University of Cambridge}
\city{Cambridge}, 
\country{United Kingdom}}

\affil[PM]{%
\orgdiv{Osservatorio Astronomico di Trieste},
\orgname{INAF},
\orgaddress{%
\city{Trieste},
\postcode{I-34143},
\country{Italy}}}

\affil[AMa]{%
\orgdiv{Dipartimento di Fisica e Astronomia}, 
\orgname{Universit\`{a} degli Studi di Firenze}, 
\orgaddress{%
\city{Sesto Fiorentino, Firenze}, 
\postcode{I-50019}, 
\country{Italy}}}

\affil[JRM]{%
\orgdiv{Departamento de F\'isica Te\'orica e Experimental, Universidade Federal do Rio Grande do Norte, Campus Universit\'ario}
\city{Natal}, 
\postcode{59078-970}, 
\country{Brasil}}

\affil[AM1]{%
\orgdiv{Dipartimento  di  Fisica  e  Astronomia  “Augusto  Righi”}, 
\orgname{Alma  Mater Studiorum, Universit\`a  di Bologna}, 
\orgaddress{%
\city{Bologna}, 
\postcode{40129}, 
\country{Italy}}}

\affil[AM2, LO, DR]{%
\orgdiv{Osservatorio di Astrofisica e Scienza dello Spazio}, 
\orgname{INAF}, 
\orgaddress{%
\city{Bologna}, 
\postcode{40129}, 
\country{Italy}}}

\affil[NNa]{%
\orgdiv{Department}, 
\orgname{Organization}, 
\orgaddress{%
\city{City}, 
\postcode{100190}, 
\state{State}, 
\country{Country}}}

\affil[CP]{%
\orgdiv{European Southern Observatory},
\orgaddress{%
\city{Garching bei M\"{u}nchen},
\postcode{85748},
\country{Germany}}}

\affil[KP1]{%
\orgdiv{Leibniz Institute for Astrophysics Potsdam (AIP)}, 
\orgaddress{%
\city{Potsdam}, 
\postcode{14482}, 
\country{Germany}}}

\affil[KP2]{%
\orgdiv{Potsdam University, Institute for Physics and Astronomy}, 
\orgaddress{%
\city{Potsdam-Golm}, 
\postcode{14476}, 
\country{Germany}}}

\affil[SN]{%
\orgdiv{Università degli Studi di Firenze}, 
\orgname{Dipartimento di Fisica e Astronomia}, 
\orgaddress{%
\city{Firenze}, 
\postcode{50125}, 
\country{Italy}}}

\affil[PT]{%
\orgdiv{Sorbonne Universit\'es, UPMC Univ Paris 6 et CNRS, UMR 7095, Institut d'Astrophysique de Paris}
\city{Paris}, 
\country{France}}

\affil[KV1]{%
\orgdiv{Department of Physics and Astronomy, University of Victoria}
\orgaddress{%
PO Box 1700, STN CSC}
\city{Victoria},
\state{BC}, 
\postcode{V8W 2Y2}, 
\country{Canada}}

\affil[GAW]{%
\orgdiv{Department of Physics and Space Science}, 
\orgaddress{PO Box 17000, Station Forces
\city{Kingston, Ontario}, 
\postcode{K7K 7B4}, 
\country{Canada}}}

\affil[AZ]{%
\orgdiv{Osservatorio Astronomico di Brera}, 
\orgname{INAF}, 
\orgaddress{%
\city{Merate}, 
\country{Italy}}}

\abstract{%
The ArmazoNes high Dispersion Echelle Spectrograph (ANDES)
is the optical and near-infrared high-resolution echelle spectrograph
envisioned 
for the European Extremely Large Telescope (ELT).~
We present a selection of science cases, supported by new calculations and simulations, where ANDES could 
enable major advances in the fields of stars and stellar populations.
We focus on three key areas, including
the physics of stellar atmospheres, structure, and evolution;
stars of the Milky Way, Local Group, and beyond; and 
the star-planet connection.
The key features of ANDES are its wide wavelength coverage at high spectral resolution and its access to the large collecting area of the ELT.~
These features position ANDES to address the most compelling 
and potentially transformative
science questions in stellar astrophysics of the decades ahead,
including questions which cannot be anticipated today.
}

\keywords{%
Star clusters (1567),
Stellar atmospheres (1584),
Stellar evolution (1599),
Stellar physics (1621),
Stellar populations (1622),
High resolution spectroscopy (2096),
Galactic archaeology (2178)}

\maketitle

\section{Introduction}
\label{sec:intro}

%

The European Extremely Large Telescope (ELT) will be the world's largest optical and near-infrared (NIR) telescope once it is commissioned, and it will retain that status for the foreseeable future.
Design work is underway for the ArmazoNes high Dispersion Echelle Spectrograph (ANDES), which will be the workhorse optical and NIR high-resolution spectrograph for the ELT.~
ANDES is poised to capitalize on major revolutions that will occur in the field of stellar astrophysics in the decades ahead.
These revolutions are enabled by new views of the infrared sky from JWST, new cadences of the changing optical sky from the Rubin Observatory, new enormous multiplexing capabilities of instruments such as the Multi Object Optical and Near-infrared Spectrograph (MOONS) and 4-meter Multi-Object Spectroscopic Telescope (4MOST), and other observational facilities.
ANDES will be the right instrument to build on these discoveries by enabling new stellar physics experiments and probing stellar populations in new environments in the 2030s and beyond.

Studies on the physics of stars, their interactions with exoplanets, and the stellar populations of the Milky Way and external galaxies benefit enormously from general-purpose high-resolution spectrographs that cover a broad wavelength coverage.
High-resolution ultraviolet (UV), optical, and NIR spectrographs routinely serve a wide range of applications across all major fields of astrophysics.
Such instruments on UV-optical-NIR telescopes have led to a steady stream of discoveries for generations. 

The baseline design for ANDES includes three fiber-fed spectrographs that simultaneously record wavelengths, $\lambda$, from roughly 0.40 to 1.80~$\mu$m at a spectral resolving power, $R \equiv \lambda/\Delta\lambda$, of 100,000.
Design goals include extensions to the $U$ (0.35--0.41~$\mu$m) and $K$ (1.80--2.40~$\mu$m) bands, where the $K$ channel is contained within a separate spectrograph, as well as an intermediate spectral resolving power of $R$ = 50,000.
The ANDES design includes a two-fiber seeing-limited observing mode (fiber diameter = 0.76'' on sky) and a 61-\mbox{spaxel} fiber matrix mode (sizes ranging from 5 to 100 mas/\mbox{spaxel}) that functions like an integral field unit (IFU) fed by a single conjugate adaptive optics (SCAO) system for $\lambda >$ 0.95~$\mu$m.

The design and capabilities of ANDES are described in greater detail in \citet{marconi22}.
The large collecting area of the ELT and the capable, general-purpose ANDES design will play a transformative role in shaping stellar astrophysics in the coming decades.

In this paper, we highlight key open questions in the field of stellar astrophysics today where ANDES is poised to enable major advances.
We focus on three key areas:\ 
the physics of stellar atmospheres, structure, and evolution (Section~\ref{sec:stellarphysics});
individual stars and resolved stellar populations in the Milky Way, Local Group, and beyond (Section~\ref{sec:milkyway}), and
the connections between stars and their planetary systems (Section~\ref{sec:starplanet}).
Other papers in this series focus on the transformative role that ANDES may play in advancing our understanding of 
exoplanets and protoplanetary disks,
galaxies and the intergalactic medium, and
cosmology and fundamental physics.
We recognize that the most compelling science cases of tomorrow may differ from the ones presented here.
Yet these science cases underscore the sustained demand from the community for a capable high-resolution spectrograph on the ELT with broad wavelength coverage,
and the enormous discovery potential that ANDES will enable. All the calculations made in the following sections have been made with the latest version of the ANDES ETC\footnote{\href{https://andes.inaf.it/instrument/exposure-time-calculator/}{https://andes.inaf.it/instrument/exposure-time-calculator/}} (v1.1).

\section{The Physics of Stellar Atmospheres, Structure, and Evolution}
\label{sec:stellarphysics}

\subsection{Asteroseismology of stellar populations across the Milky Way and beyond}
\label{sec:asteroseismology}

Asteroseismology is the study of stellar oscillations.
It can reveal information about the interiors of stars that is otherwise inaccessible.
It also enables accurate stellar ages to be determined, providing a new dimension of our view of the formation of our Galaxy and its accretion history \citep[e.g.,][]{2020NatAs...4..382C,2021NatAs...5..640M}.  
Previous asteroseismic studies performed with photometric space missions, such as CoRoT, Kepler, and TESS \citep[][and references therein]{garcia2019, 2022ARA&A..60...31K} 
were only able to observe relatively bright nearby main sequence stars or red giants \citep[e.g.,][]{2019A&A...625A..33C}. These observations have been limited to a small fraction of the stellar populations found throughout the Milky Way \citep[e.g.,][]{2021A&A...645A..85M, 2023MNRAS.tmp.1842S}.

ANDES will be able to detect and measure small ($\sim$~10--30~cm/s) Doppler shifts of spectral lines.
These oscillations offer the promise of being able to determine ages accurately to $\approx$~1--2~Gyr for large numbers of individual field stars, which also has not been possible previously. ANDES will also enable measurements of oscillations in much more distant main sequence stars in the disk of our galaxy, faint halo stars, and red giants in Local Group galaxies for the first time. The observed eigenmodes can be used to determine the age of individual extragalactic stars,
providing a unique opportunity to date components of Local Group galaxies for the first time and provide new constraints on galaxy formation mechanisms.

At the other end of the Hertzsprung-Russell (HR) diagram, asteroseismology with ANDES will be pivotal in our understanding of the fundamental physical parameters of M dwarfs, in particular mass and radius.
%
Thermodynamic and Solar-like pulsations have been predicted for young and main sequence M dwarfs \citep{RodriguezLopez2012, RodriguezLopez2014}, but they have not yet been detected (see \citealt{RodriguezLopez2019} for a review).
The predicted pulsation periods are short, on the order of tens of minutes, and the expected 
radial velocity amplitudes of oscillations are small, below 1~m~s$^{-1}$.
The intrinsic faintness of M dwarfs (Gaia $G_{\rm BP} \gtrsim 10.0$ mag) presents an observational challenge that ANDES can overcome, because it will be able to obtain time-resolved spectroscopic data of M dwarfs. These measurements of line profiles and radial velocity variations at a consistently high cadence and precision are necessary to discover and characterize the expected pulsation modes of M dwarfs.

\subsection{Backbone measurements of stellar activity and magnetism}\label{sec:activity}

In the context of late-type stars, stellar activity can be broadly summarized as any observed manifestation of their magnetism. It underlies many of the complex behaviors in those stars, including the relationships between internal processes, such as convection and dynamos, with external ones, such as stellar winds and angular momentum loss \cite[][]{2008LRSP....5....2H, 2021LRSP...18....3V, 2023SSRv..219...70I}. The stellar activity must be characterized when searching for exoplanets.  It is especially critical when searching for Earth-mass planets orbiting Sun-like stars, because the lack of such information can easily lead to false positives, and it currently represents the limiting factor when pushing the radial velocity method to the required precision limit at the cm~s$^{-1}$ level \cite[][]{2013AN....334..616H,2016A&A...585A.144H,2017A&A...606A.107O,Hojjatpanah2020}. Moreover, several studies have demonstrated how activity in young stars ($t<100$\,Myr) can shape the stellar spectrum in ways we still cannot account for, affecting the atmospheric parameters and the abundances we infer with a classic spectroscopic analysis \cite[][]{2019galarza,2020spina,2020baratellaA,2021baratella}. Understanding which mechanism(s) of activity dominate such alterations is pivotal.

ANDES has the potential to be the most powerful astronomical instrument to collect data on stellar atmospheric activity and magnetic fields, because of its broad wavelength coverage and the large collecting area of the ELT.~ Table~\ref{Table:ActInd} shows a number of current and proposed future instruments, some with the capability to study stellar magnetic fields simultaneously in the whole wavelength range with the most widely used activity indicators. Compared to other facilities, ANDES excels because of its wide spectral coverage from the blue to the red (0.35--2.4~$\mu$m).  The collecting area of the ELT and spectral resolving power of ANDES will provide the completeness required to study the complex connection between stellar atmospheric activity and magnetic physics, and they will allow us to simultaneously disentangle sensitive phenomena in the blue and red parts of the spectrum. For example, magnetic features at different atmospheric layers in the blue and red respond differently to exoplanet transits \cite[see][]{2022BAAS...54e.407R,2023RASTI...2..148R}. 

\begin{table}
\begin{tabular}{l c r c c c c c}
 \hline
 \vspace{0.1cm}
Instrument (Telescope) & $\lambda$ ($\mu$m) & Res. & Ca~\textsc{ii}$_{\rm H\&K}$ & H$\alpha$ & Ca~\textsc{ii}$_{\rm IRT}$ &  B$_{\rm ZB}$ & SIM \\
\hline      
ANDES (ELT\,39\,m) [baseline]  & \textcolor{black}{0.40-1.80} &  100\,k & \textcolor{red}{\xmark} & \textcolor{green}{\cmark} & \textcolor{green}{\cmark}  & \textcolor{green}{\cmark} & \textcolor{red}{\xmark} \\
ANDES (ELT\,39\,m) [goal]   & \textcolor{black}{0.35-2.40} &  100\,k & \textcolor{green}{\cmark} & \textcolor{green}{\cmark} & \textcolor{green}{\cmark}  & \textcolor{green}{\cmark} & \textcolor{green}{\cmark} \\
CARMENES (CAHA\,3.5m) & 0.52-1.71 &  90\,k & \textcolor{red}{\xmark}   & \textcolor{red}{\xmark}   & \textcolor{green}{\cmark}  & \textcolor{green}{\cmark} & \textcolor{green}{\cmark} \\
CRIRES+ (VLT\,8.2\,m) & 0.95-5.30 & 100\,k & \textcolor{red}{\xmark} & \textcolor{red}{\xmark} & \textcolor{red}{\xmark} & \textcolor{green}{\cmark} & \textcolor{red}{\xmark} \\
ESPRESSO (VLT\,8.2\,m)    & 0.38-0.68 &  140\,k & \textcolor{green}{\cmark} & \textcolor{green}{\cmark} & \textcolor{red}{\xmark}    & \textcolor{red}{\xmark}   & \textcolor{green}{\cmark} \\
G-CLEF (GMT\,25.4\,m)  & 0.35–0.90 &  108\,k & \textcolor{green}{\cmark} & \textcolor{green}{\cmark} & \textcolor{green}{\cmark}  & \textcolor{red}{\xmark}   & \textcolor{green}{\cmark} \\ 
GIANO-B (TNG\,3.6\,m) & 0.95-2.45 & 50\,k & \textcolor{red}{\xmark} & \textcolor{red}{\xmark} & \textcolor{red}{\xmark} & \textcolor{green}{\cmark} &  \textcolor{green}{\cmark} \\
HARPS (ESO\,3.6\,m) & 0.38-0.69 &  120\,k & \textcolor{green}{\cmark} & \textcolor{green}{\cmark} & \textcolor{red}{\xmark}    & \textcolor{red}{\xmark}   & \textcolor{green}{\cmark} \\
HARPS-N (TNG\,3.6\,m) & 0.37-0.69 & 120\,k & \textcolor{green}{\cmark} & \textcolor{green}{\cmark} & \textcolor{red}{\xmark} & \textcolor{red}{\xmark} & \textcolor{green}{\cmark} \\
HIRES (Keck\,10\,m)    & 0.30-1.00 &   85\,k & \textcolor{green}{\cmark} & \textcolor{green}{\cmark} & \textcolor{green}{\cmark}  & \textcolor{green}{\cmark} & \textcolor{red}{\xmark} \\
NIRPS (ESO\,3.6\,m) & 0.95-1.80 &  100\,k & \textcolor{red}{\xmark}   & \textcolor{red}{\xmark}   & \textcolor{green}{\cmark}  & \textcolor{green}{\cmark} & \textcolor{green}{\cmark} \\
PEPSI (LBT\,2x8.4\,m)  & 0.38-0.91 &  250\,k & \textcolor{green}{\cmark} & \textcolor{green}{\cmark} & \textcolor{green}{\cmark}  & \textcolor{green}{\cmark} & \textcolor{red}{\xmark} \\
UVES (VLT\,8.2\,m) &   0.30-1.10 & 80\,k & \textcolor{green}{\cmark} & \textcolor{green}{\cmark} & \textcolor{green}{\cmark} & \textcolor{green}{\cmark} & \textcolor{red}{\xmark} \\
VISION (CFHT\,3.6\,m) & 0.37-2.44 &   75\,k & \textcolor{green}{\cmark}  & \textcolor{green}{\cmark}  & \textcolor{green}{\cmark}  & \textcolor{green}{\cmark} & \textcolor{green}{\cmark} \\
\hline  
                       \multicolumn{3}{r}{Number of ADS references:} &  256  &   130    &  70    &  75  \\    
\hline
\end{tabular}
\vspace{0.2cm}
\caption{Scope of capabilities comparison for different facilities to study the most commonly used stellar activity indicators (the Ca~\textsc{ii} H\&K lines at 393.3 and 396.8~nm, the H$\alpha$ line at 656.3~nm, and the Ca~\textsc{ii} NIR triplet (IRT) lines at 849.8, 854.2, and 866.2~nm) and magnetic field measurements via Zeeman Broadening (ZB) at $\sim$1$\mu$m. The second column lists the wavelength range covered by each instrument. The third column lists the spectral resolving power (Res.) in units of 1000 (k). The last column indicates the simultaneity (SIM) in the observable ranges. The bottom row lists the number of refereed publications in the Astrophysics Data System (ADS) as of 2023 with an explicit mention of each activity indicator. 
}\label{Table:ActInd}
\end{table} 

This potential could be only achieved through the implementation of the proposed ANDES $U$ channel extension.
The emission in the cores of collisionally dominated Ca~\textsc{ii} H\&K Fraunhofer lines at 393.3 and 396.7~nm (Ca~\textsc{ii}$_{\rm H\&K}$) constitutes the most consistent and reliable indicator of chromospheric magnetic activity observable using ground-based instruments.
This measure also provides the longest observational time baseline \citep{2014ApJ...794..144R,2022A&A...658A..16B}. Ca~\textsc{ii}$_{\rm H\&K}$ is also well correlated to the coronal activity observed in X~rays \cite[see][]{2011ApJ...743...48W,2018MNRAS.473.4326A,2019A&A...631A..45S}. Other less sensitive activity indicators, such as the Ca~\textsc{ii} infrared triplet (Ca~\textsc{ii}$_{\rm IRT}$) or H$\alpha$, have been used in the past, mainly due to the lack of Ca~\textsc{ii}$_{\rm H\&K}$ measurements. The Ca~\textsc{ii}$_{\rm IRT}$ lines can also be strongly affected by changes in T$_{\rm eff}$ and metallicity as shown by analyses of model spectra \cite[see][]{2005A&A...430..669A,2017A&A...605A.113M}. In solar observations, the prominent Balmer series lines
respond mostly to the photosphere, exhibiting a broad scatter that correlates with sunspot transits but not chromospherically sensitive lines \cite[see][]{2021A&A...646A..81M,2023ApJ...951..151C}. Similarly, a strong scatter from H$\alpha$ is observed in Sun-like stars when compared with Ca~\textsc{ii}$_{\rm H\&K}$ lines and X~rays \cite[see][]{,2023MNRAS.524.5725A}. Therefore, the proposed $U$ channel extension to the baseline design of ANDES would enable these exciting new measurements and robustly characterize stellar activity for the different atmospheric layers across spectral types.
It would enable the new measurements of phenomena spanning timescales of seconds to decades, to be benchmarked alongside 
legacy data from other facilities. It is no coincidence that this activity indicator has had the largest impact on different sub-fields of astrophysics, as shown by the number of references listed in Table~\ref{Table:ActInd}.


ANDES will also offer the opportunity to study stellar magnetism directly in unprecedented detail for stars all across the HR diagram. Among cool stars of spectral types F to M, Zeeman Broadening (ZB) measurements use Ti~\textsc{i} and FeH lines around $\lambda \approx$ 1~$\mu$m \citep{2019A&A...626A..86S,2022A&A...662A..41R}, while Zeeman Intensification (ZI) measurements use a set of Fe~\textsc{i} and Ti~\textsc{i} lines spread throughout the $\lambda \approx$ 0.42--0.85~$\mu$m range \citep{2017ApJ...835L...4K,2020A&A...635A.142K}. Only $\lesssim 400$ of those stars have measurements of their magnetic fields via ZB or ZI methods, yet the expectation is that all such stars host dynamo-driven magnetic fields of varying levels of intensity. With magnetic fields being the main driver of the space weather environment of cool stars (i.e. high-energy radiation, stellar winds, energetic transient events; see \citealt{2019BAAS...51c.113D}), any exoplanet-related observing programme on ANDES will allow direct measurements of the stellar magnetic field. Apart from the information on the field itself, this will enable a more realistic characterization of the exoplanet habitability conditions, by placing constrains for models investigating processes such as atmospheric escape and photon- and particle-driven atmospheric chemistry \citep[see e.g.][]{2020IJAsB..19..136A}.

Magnetic fields are also crucial actors affecting the observational properties and evolution of intermediate- and high-mass stars \citep[e.g.][]{2009ARA&A..47..333D}. Those magnetic fields are broadly considered to be ``fossil fields,'' which are slowly decaying remnants from flux advection, mass transfer, or mergers at earlier phases of evolution. The presence of magnetic fields in intermediate-mass main sequence stars influences atmospheric microscopic diffusion processes, leading to overall peculiar chemistry and nonuniform vertical and lateral distributions of chemical abundances. Magnetic fields channel and confine stellar winds in the hottest and most massive main sequence stars of spectral types B and O, leading to efficient shedding of rotational angular momentum and mass-loss quenching \citep{2013MNRAS.429..398P}. These effects fundamentally impact the evolution of the star, including its subsequent supernova characteristics and remnant properties \citep{2017MNRAS.466.1052P,2019MNRAS.485.5843K}. Studies of the magnetic properties of such stars, based on resolving the wavelength separations of magnetically split lines, have mostly been conducted using optical wavelengths \citep[e.g.][]{Mathys2017,Hubrig2019}.
Only recently has \citet{magsplit} conducted the first serious exploitation of this phenomenon using the $H$ band.

ANDES will be well-positioned to revolutionize this field, because the majority of the current data on stellar magnetic fields in both low- and high-mass stars has been obtained with instrumentation on $\leq4$~m class telescopes.
ANDES will be able to distinguish subtle line profile differences among intermediate and massive stars on the main sequence.
ANDES will also be superior to existing facilities in its ability to better resolve lines in more rapidly rotating stars for a given magnetic field strength, and resolve Zeeman-split components in stars with weaker magnetic fields for a given Doppler line width. 

Finally,
ANDES will also provide an unprecedented opportunity to understand how stellar activity and magnetism vary as a function of stellar type and age by performing studies of stars in open clusters. The overwhelming majority of previous activity and magnetic field measurements have been made for field stars, where trends with other parameters are nearly impossible to discern. The large collecting area of the ELT will enable ANDES to study intrinsically fainter stars at larger distances, potentially revealing how stellar mass, age, and environmental aspects influence the behavior of stellar activity and magnetism, with implications for multiple fields of modern astrophysics.

\subsection{Gravity darkening:\ a new measure of stellar rotation}
\label{sec:gravitydarkening}

Rotation is a fundamental parameter that governs the physical structure and evolution of stars.
For example, rotation induces internal mixing, which in turn impacts the stellar lifetime and the surface chemical abundances. 
Two important consequences of stellar rotation for the stellar photosphere are geometrical distortion (flattening) and gravity darkening.
Both effects can influence the observed stellar flux and emergent spectrum because flattening induces a local effective gravity that is stronger at the poles than at the equator. 
Similarly, gravity darkening induces a latitudinal dependence on the photospheric flux emission, with a local effective temperature higher at the poles than at the equator.  
This latitudinal dependence of local effective gravity and temperature breaks the spherical symmetry of a fast-rotating star, so the observed spectrum reveals the viewing inclination angle, $i$.
Once $i$ is known, more fundamental physical parameters can be derived, including the true rotation velocity and mass \citep{2023arXiv230700082L}.
These quantities are intimately linked to the internal circulation of matter, energy, and angular momentum, as well as angular momentum loss and non-spherical mass loss \citep{2000A&A...361..159M,2019A&A...625A..88G,2019A&A...625A..89G}.

High spectral resolution and ultra-high signal-to-noise (S/N $> 1000$) data enable direct measurements of these quantities, which
are directly encoded in the continuum and lines of a spectrum of a rotating star.
Spectra of this quality is not generally obtainable today, except for the very brightest stars in the sky
\citep{2020MNRAS.499.1126T,2021MNRAS.505.1905T}.
ANDES will enable new measurements of these quantities for fast-rotating stars, because of its high spectral resolution, wide wavelength coverage spanning the optical to the NIR regimes, and the large collecting area of the ELT.~
Gravity darkening and rotation velocities measured by ANDES would be able to provide novel constraints on models of stellar structure and evolution.
Measuring gravity darkening in many fast-rotating stars across the HR diagram has the potential to provide critical constraints that lead to a more profound understanding of stellar physics.

\subsection{Lithium in stars: primordial versus  evolutionary and environmental effects}
\label{sec:lithium}



Hydrogen, helium  and lithium are produced in the first few minutes after the Big Bang.
Reconciling predictions for the primordial lithium abundance with observations remains one of the major open questions in astrophysics today.
The primordial lithium abundance can be predicted from standard Big Bang nucleosynthesis, where it emerges as a function of the cosmological baryonic density, $\Omega_{B}$, a parameter that quantifies the amount of ordinary matter. 
$\Omega_{B}$ can be derived to high precision from observations of the cosmic microwave background \citep[e.g.,][]{coc13}.
This predicted abundance is in conflict with the lithium abundances observed in the oldest stars.
The vast majority of unevolved Population~II (Pop~II) dwarf stars in the Milky Way with metallicities, [Fe/H], between $\sim -3.0$ and $\sim -1.0$  share a similar surface lithium abundance, the so-called Spite Plateau value \citep{Spite1982}.
This derived lithium abundance is a factor of 3--4 lower than that predicted by the standard Big Bang nucleosynthesis model.

\begin{figure}[h]%
\centering
\includegraphics[width=0.8\textwidth]{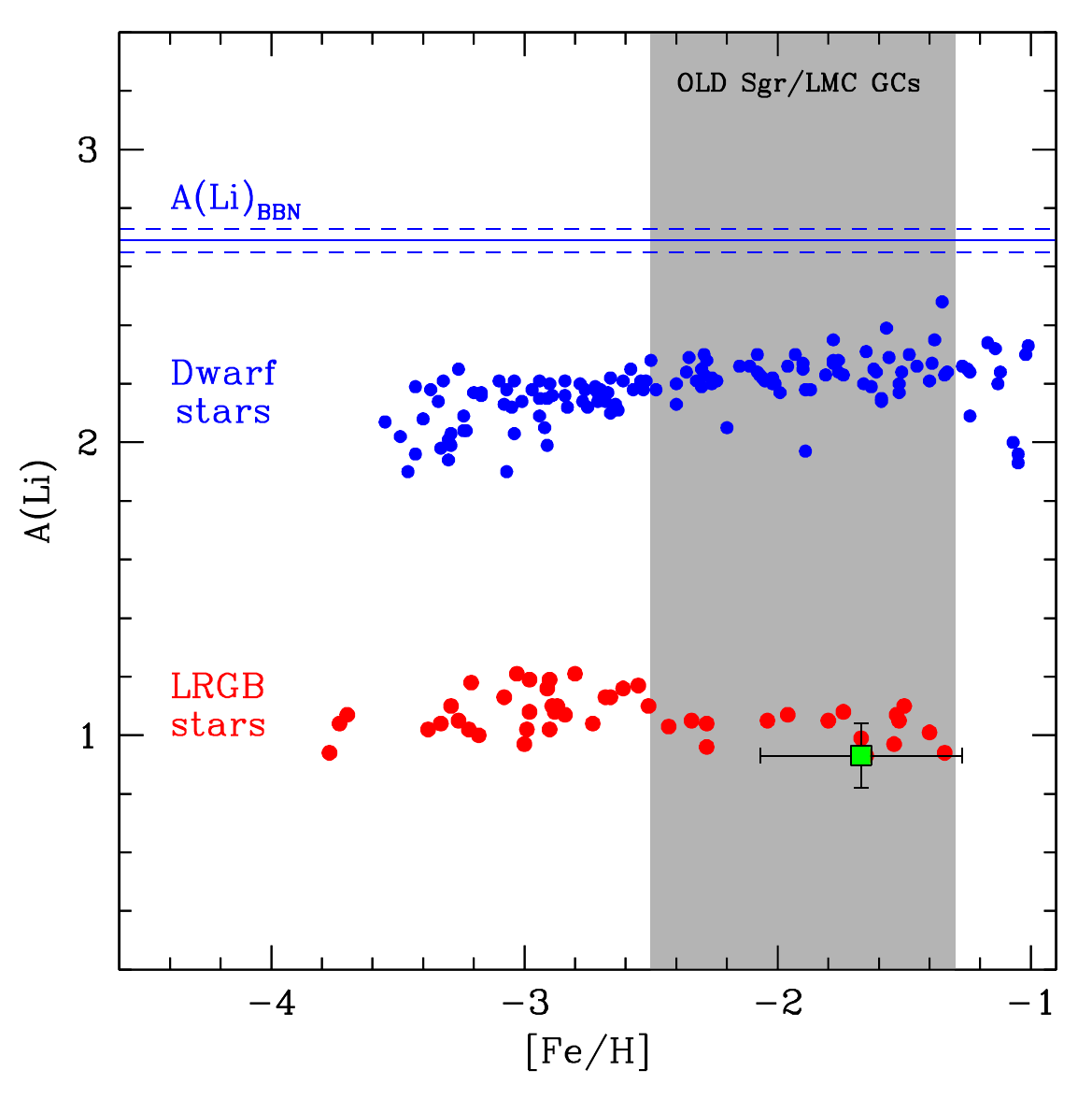}
\caption{
Surface lithium (Li) abundance as a function of [Fe/H] for Milky Way dwarf stars \citep[blue circles][]{bonifacio97,asplund06,aoki09,hosford09,melendez10} and stars on the lower red-giant branch \citep[red circles][]{Mucciarelli2022}. 
The green square marks the lithium abundance derived from the stacked spectrum of 51 lower RGB stars in the Sagittarius globular cluster M54 \citep{Mucciarelli2014}.
The blue solid line denotes the lithium abundance (and $\pm 1\sigma$ uncertainties) inferred from Big Bang nucleosynthesis \citep{coc13}.
The grey shaded area indicates the range of [Fe/H] of the old Sagittarius and Large Magellanic Cloud globular clusters that are ideal targets for ANDES.}
\label{fig:lithium}
\end{figure}

A complementary view of the initial A(Li) of Pop~II stars relies on stars on the lower red giant branch \citep{Mucciarelli2012}. 
These giants are characterised by a constant Li abundance, drawing a {\sl Plateau} that mirrors the A(Li) {\sl Plateau} observed  with the dwarf stars but at a lower abundance (A(Li)$\sim$0.9 dex) as shown  in Figure~\ref{fig:lithium}. The lower Li content is due to the first dredge-up episode and 
the increasing size of the convective envelope. 
Stellar models including diffusion and  additional mixing are able to reconcile the measured A(Li) in both dwarf and giant stars with the cosmological value \citep{Mucciarelli2022}.
The key upside of this approach is that these stars are about 4~magnitudes brighter than the main sequence turnoff at the wavelength of the Li~\textsc{i} line (at 670.7~nm), enabling to investigate A(Li) in stars more distant than those usually observed with the dwarf stars. 

ANDES will be able to collect high-resolution spectra of the Li~\textsc{i} line in individual lower red-giant branch stars in extragalactic systems---including the Magellanic Clouds and the Sagittarius dwarf galaxy---for the first time.
Lithium abundances in these stars could be derived with a precision of 0.06 (0.03) dex from ANDES spectra with S/N = 50 (100) per pixel.
ANDES will be capable of collecting spectra of this quality for stars in these Local Group galaxies in about 1 (5) hr of observing time for $V$ = 18.5, and around 2 (8) hr for $V$ = 19.0. This new capability will enable investigations into whether the lithium abundance discrepancy exists in extragalactic systems, which experienced different star formation histories and whose stars exhibit different chemical abundance patterns.
ANDES will thus have the potential to establish the environmental dependence of the lithium abundance discrepancy that continues to vex the cosmological and stellar astrophysics communities.


\subsection{Atomic diffusion in low-mass stars}

Atomic diffusion is the slow transport of chemicals due to temperature, pressure and abundance gradients 
occurring in the radiative stellar layers. Specifically, pressure and temperature gradients push the ions toward the center, decreasing the surface abundances (gravitational settling), while the interaction with the radiation field and the concentration
gradients lead to the opposite effect (radiative levitation). The surface abundance variations along the evolutionary sequence on the interplay of these two competitive processes,
which are maximized in stars around the main sequence turnoff point \citep[e.g.,][]{1984ApJ...282..206M, salaris17}. Subsequent evolution along the red giant branch restores the initial abundances, as the deepening outer convection zone mixes elements back to the surface.
%
Observations show that the abundance differences between dwarf and giant stars are significantly smaller than those expected from stellar models that include atomic diffusion \citep{Gratton2001,Cohen05,Korn2006,Mucciarelli11}, indicating that diffusion is strongly inhibited by additional mixing processes (on top of convection) that operate in main sequence stars. The nature of this additional mixing is still unknown, but its correct characterization is crucial to better constrain stellar evolution models including these processes, understand potential biases affecting the chemical abundances of unevolved field stars, and
precisely quantify the cosmological lithium problem (Section~\ref{sec:lithium}; \citealt{Korn2006,Gavel2021}). At the very least, the joint analysis of dwarfs and giants in Galactic
archaeology requires an empirical understanding of the surface abundance
corrections caused by atomic diffusion.

The efficiency of the atomic diffusion and the unidentified source of additional mixing along the main sequence depend on the stellar mass and temperature, so precise chemical abundances of different elements at different masses are crucial to constrain these effects.
Globular clusters offer a unique opportunity to investigate these effects, by comparing abundances in relatively unevolved stars (turnoff point, subgiants) with more evolved 
stars (red giants).
A handful of globular clusters and a couple of open clusters have previously been studied in search for element-specific diffusion features,
such as for Li
\citep{Korn2007,Lind2009,Nordlander2012}. 
The main limitation in the spectroscopic studies performed so far has been the faintness of unevolved stars 
in Milky Way globular clusters, where magnitudes at the main sequence turnoff are $V \approx$~16.5--19.0, so high-resolution spectroscopic observations are time-consuming to
prohibitive. To this day, the main sequences of globular clusters remain
unexplored at high spectral resolution. 

\begin{figure}[h]%
\centering
\includegraphics[width=1\textwidth]{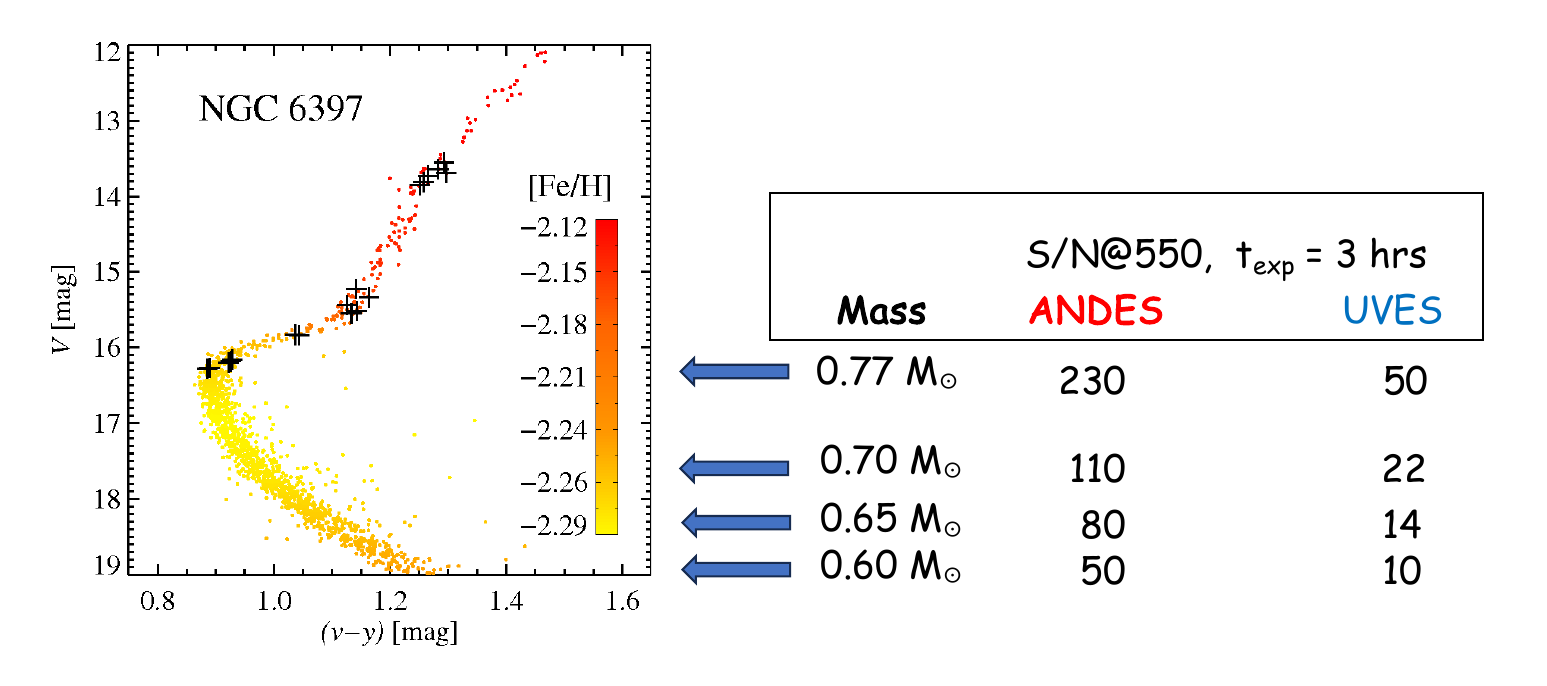}
\caption{ 
Color-magnitude diagram of the nearby globular cluster \mbox{NGC~6397}, color-coded according to [Fe/H] abundance ratios predicted by a theoretical model of atomic diffusion
calibrated according to the results of \citet{Korn2007}.
Stellar masses along the main sequence are labeled, together with the expected S/N ratios at 550~nm with three hours of exposure time using ANDES and UVES.} 
\label{diffusion}
\end{figure}

ANDES will provide an unprecedented opportunity to study atomic diffusion by deriving precise chemical abundances of Li, Fe, and other metals
in unevolved and evolved stars of globular clusters.
In the closest clusters, such as NGC~6397
(Figure~\ref{diffusion}),
ANDES will be able to collect spectra of stellar masses as low as $\sim$~0.6~$M_{\odot}$, which are not obtainable using any other existing high-resolution spectrograph.

\subsection{Constraining the initial mass function using CNO isotopes in unevolved stars}
\label{sec:cnoisotopes}

Determining the shape of the initial mass function (IMF) and its dependence on the physical conditions of the medium out of which the stars form is a major open question of modern astrophysics.
Isotopic abundance ratios provide a powerful diagnostic of the IMF in systems where direct measurements, such as star counts, are impossible. 
Stars of different masses produce different isotopic ratios among carbon, nitrogen, and oxygen, which are collectively referred to as CNO elements.  Measuring isotopic ratios among the CNO elements provides a critical test of the possible non-universality of the IMF \citep{henk1993,papa2014,roma2017}. 
Unevolved stars whose abundances have not been altered by internal evolutionary processes hold the key to studying the chemical evolution of rare isotopes of CNO in the Milky Way.
The current sample sizes are far too small to provide meaningful constraints on the IMF.~
The $^{12}$C/$^{13}$C and $^{16}$O/$^{18}$O ratios have been measured only in a few nearby objects \citep[see][for a recent review]{roma2022} and it is unclear whether the $^{14}$N/$^{15}$N ratio can be measured at all in unevolved stars. 

\begin{figure}[h]%
\centering
\includegraphics[width=0.7\textwidth]{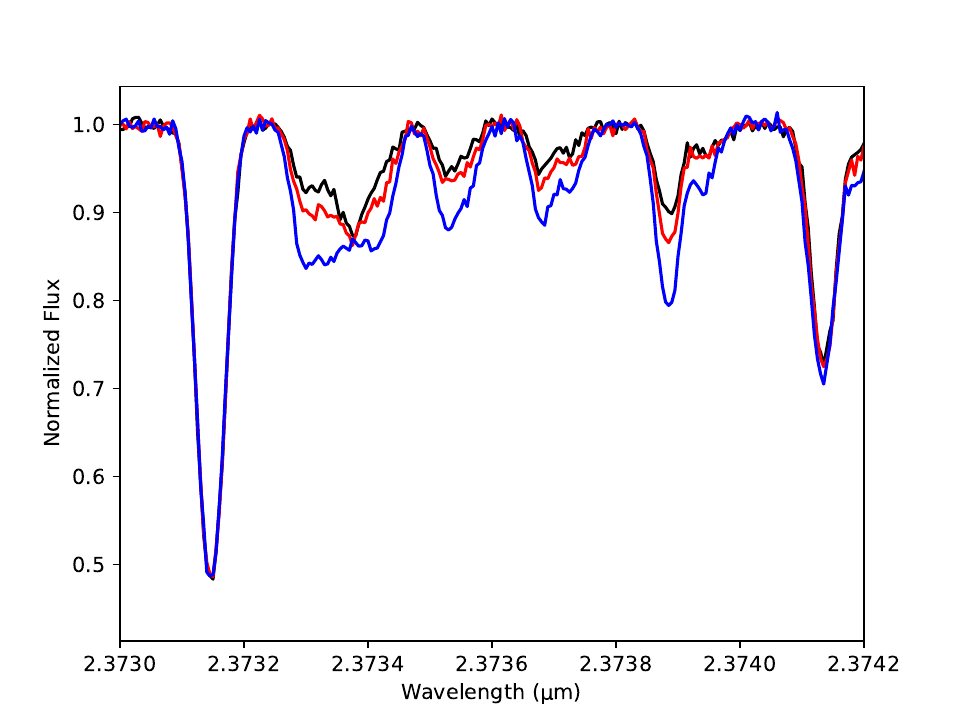}
\caption{Synthetic spectra in the $K$ band for a fully-convective M~dwarf star ($T_{\mathrm{eff}} =$ 3000~K, log\,g~= 4.5~dex, [M/H]~= 0) observed with ANDES ($R =$ 100,000) at S/N~= 200. The blue, red, and black curves refer to different isotopic ratios ($^{12}$C/$^{13}$C~= 89, 60, and 30, respectively).
The CO features that enable these measurements are detectable in several wavelength ranges, in addition to this one, across the $K$ band.} 
\label{fig:12C13C}
\end{figure}

If ANDES includes a $K$ channel, this capability would enable measurements of the $^{12}$C/$^{13}$C ratios in a large sample of unevolved stars for the first time (see Fig.~\ref{fig:12C13C}). According to ANDES ETC, one hour of science exposure shall suffice to reach S/N~$\simeq$ 200, sufficient to measure the faintest $^{13}$CO features at a few hundredths of the continuum in the spectrum of a fully convective M dwarf as faint as $K \simeq 13.5$ (in Vega system), which corresponds to a distance of about 200~pc (for comparison, CRIRES+ mounted on UT3 resolves the $^{13}$CO features of such an object only up to $\sim$50~pc distance). Longer integration times of, say, 3 hours, will allow to reach $d =$ 275~pc, thus greatly expanding the sample of potential targets unobservable at high spectral resolution using current facilities. Although the probed metallicity range will still be limited to nearly solar metallicities, it is worth noticing that ANDES will permit the determination of the pristine C isotopic ratio in a significant number of open clusters. A query to the \citet{hunt2023} catalogue returns that faint $^{13}$CO features can be measured in M dwarf spectra of 1, 25, and $\sim$60 open clusters with log(Age)~$>$ 8.2 yr within, respectively, 60, 200, and 275 pc distance from us. Determinations of the pristine $^{12}$C/$^{13}$C ratio in open clusters have implications for both stellar evolution (e.g., investigation of extra-mixing processes) and Galactic chemical evolution studies (existence and extent of inhomogeneities, if any, informing on the most likely stellar polluters).

\subsection{Nucleosynthesis of the heaviest elements:\ the rapid neutron-capture process}
\label{sec:rprocess}

Understanding the origins of the elements all across the periodic table is one of the major challenges of modern astrophysics.
The rapid neutron-capture process, or \rpro, is one of the fundamental ways that stars and stellar remnants produce the elements listed along the bottom two-thirds of the periodic table \citep{cowan21,2023Arcones}, including the actinides, transuranic elements, and potentially superheavy elements \citep{2019Eichler,holmbeck23}.
Merging pairs of neutron stars have been confirmed as one site of \rpro\ nucleosynthesis, through the spectacular detection of both electromagnetic radiation and gravitational waves from a relatively nearby event observed in 2017 (e.g., \citealt{abbott17prl,drout17,watson19}).
Yet key open questions remain, including how many additional \rpro\ sites are required by observations, which elements are made by the \rpro, and in what amounts, and what are the specific conditions required---or range of conditions permissible---for \rpro\ nucleosynthesis to occur \citep{2007Francois,2012Hansen}.

ANDES will advance our understanding of these questions because it will be highly efficient at collecting high-resolution, high-S/N optical spectra of low-metallicity \rpro-enhanced stars identified from photometric or medium-resolution spectroscopic surveys.
Many of these stars are located far beyond the Solar neighborhood \citep{aguado21,matsuno21}.
ANDES will also expand the range of potential stellar samples in Milky Way dwarf galaxies beyond the handful of brightest red giants that are marginally accessible using current facilities \citep{reichert21fnx,ji23ret2}.
Key absorption lines in late-type stars are found throughout the $U$, $B$, and $V$ channels \citep[e.g.,][]{roederer22}, including
Sr~\textsc{ii} at 407.7~nm, 
Y~\textsc{ii} at 360.0~nm, 
Zr~\textsc{ii} at 414.9~nm, 
Mo~\textsc{i} at 386.4~nm,
Ru~\textsc{i} at 379.9~nm, 
Ba~\textsc{ii} at 455.4~nm, 
La~\textsc{ii} at 398.8~nm, 
Eu~\textsc{ii} at 381.9~nm, 
Dy~\textsc{ii} at 394.4~nm, 
Yb~\textsc{ii} at 369.4~nm, 
Ir~\textsc{i} at 380.0~nm, 
Pb~\textsc{i} at 405.7~nm, 
Th~\textsc{ii} at 401.9~nm, and 
U~\textsc{ii} at 385.9~nm.

\subsection{Rare phases of stellar evolution}
\label{sec:rarehdc}

The merger of two white dwarfs (WDs) is a rare astrophysical phenomenon that concludes the life of a close stellar binary system. 
Many WD mergers will lead to Type~Ia supernova events, 
yet rare classes of WDs offer novel opportunities to explore uncommon and short-lived states of evolved stars.
The fusion of two massive (M$_{\rm total}>$~1.44 M$_{\odot}$) carbon-oxygen WDs, for example, may generate a Type~Ia thermonuclear supernova explosion
\citep{2010ApJ...714L..52S} or create a neutron star after a gravitational collapse.
Understanding the relationships between these various classes of stars, and the physics that drives them, is critical to improving population synthesis models that are widely employed across many fields of astronomy.

The hydrogen-deficient carbon (HdC) stars \citep{1984ApJ...277..355W} are supergiants with atmospheres that are composed almost entirely of helium, and they
are strong candidates to be the descendants of mergers of 
helium and carbon-oxygen WDs.~
HdC stars are rare objects.
Only about 150 are known in the Milky Way \citep{2020A&A...635A..14T,2022A&A...667A..83T}, and about 30 are known in the Magellanic Clouds \citep{2009A&A...501..985T}.
Population synthesis models predict the existence of two ensembles of HdC stars \citep{2023arXiv230910148T},
which emerge $\approx$~1 and $\gtrsim$~5~Gyr after star formation events.
Abundance studies strongly support a double-degenerate scenario for their origin \citep{2011MNRAS.414.3599J},
yet none of them have ever been detected in a binary system. 

ANDES will open a new window to study HdC stars in the Milky Way thin disk, bulge, halo, and other galaxies.
Observations of HdC stars require a high spectral resolution to obtain reliable abundances. 
HdC stars have very peculiar and varied abundances for many elements, as revealed by lines of, e.g.,  Li~\textsc{i} (670.7~nm), 
F~\textsc{i} (690.3, 739.9, and 775.5~nm), 
Sr~\textsc{ii} (407.7 and 421.5~nm), 
Ba~\textsc{ii} (455.4 and 649.6~nm), and others.
Wide wavelength coverage from the $B$ to $I$ channels is necessary to cover this set of lines, and the $U$ channel would also be beneficial.
In addition, the $^{18}$O/$^{16}$O is an important diagnostic that can be measured using CO bands observable in the $K$ channel.
%
The large collecting area of the ELT enables these evolved giants to be observed even in the Magellanic Clouds in only $\approx10$--30~min.
ANDES will be capable of generating unprecedented samples of HdC stars in our Galaxy and beyond.
This potential advance represents one of the many new windows into the exotic phases of stellar evolution that will be enabled by ANDES.~

\section{Stars of the Milky Way, Local Group, and Beyond}
\label{sec:milkyway}

\subsection{The first stars}
\label{sec:firststars}


The formation of the first stars signaled the end of the cosmic Dark Ages.
The first stars, which are often referred to as Population~III (Pop~III) stars, formed from metal-free gas at redshift $z >$~15--30 \citep[e.g.,][]{bromm2011,klessen2023}.
These first stars are predicted to have been relatively massive, with characteristic masses $\geq$~10~\msun\ \citep[e.g.,][]{hirano2014,susa2014,pagnini2023} and most likely $\leq$~100~\msun \citep{Koutsouridou2023}.
They produced the first ionizing photons, the first supernovae, the first metals, and the first stellar-mass black holes.
The first stars played a central role in shaping the nature of the first galaxies and the earliest metal enrichment.
Despite their significance, these objects have not been observed directly. 
ANDES will have the ability to reveal the nature and end states of the first stars in two important ways.
First, ANDES will be able to search for long-lived, low-mass metal-free Pop~III stars, should any exist today.
The key observational signature of these stars would be the presence of H Balmer lines in absorption and the absence of absorption lines of any metals or molecules, including 
the CH G band near 430~nm,
the Na~\textsc{i} D doublet near 589~nm,
the Mg~\textsc{i} b triplet near 518~nm,
the Ca~\textsc{ii} H \& K doublet at 393.3~nm and 396.7~nm, 
and Fe~\textsc{i} lines at 358.1~nm and 382.0~nm.

ANDES would be able to detect one or more Pop~III stars---%
or exclude their existence at high confidence---%
by collecting spectra of several hundred candidate metal-free stars with initial metallicity estimates [Fe/H]~$< -4$ \citep{hartwig15,roederer19baas}.
These data would enable new, stronger constraints on the characteristic mass and low-mass end of the Pop~III IMF for the first time \citep{rossi2021}.

Secondly,
ANDES will be well-suited to characterize the surviving descendants of extreme Pop~III stars
by determining the chemical compositions of hundreds of second-generation stars born from the remnants of Pop~III stars. 
Those chemical abundance patterns could be interpreted using new state-of-the-art models to constrain the masses and explosion energies of the elusive Pop~III supernovae themselves \citep{rossi2023,Koutsouridou2023,vanni2023a}.
The predicted end states of these extreme Pop~III stars include 
energetic pair-instability supernovae \citep[PISNe;][]{HW2002,salvadori19,aguado2023,xing23};
high-energy hypernovae \citep{placco2021,skuladottir21}; and
low-energy faint supernovae that led to the formation of stars highly enhanced in carbon and deficient in iron, the so-called carbon-enhanced metal-poor stars \citep[e.g.,][]{Iwamoto2005,debennassuti17,rossi2023}.

\begin{figure}[h]%
\centering
\includegraphics[width=0.95\textwidth]{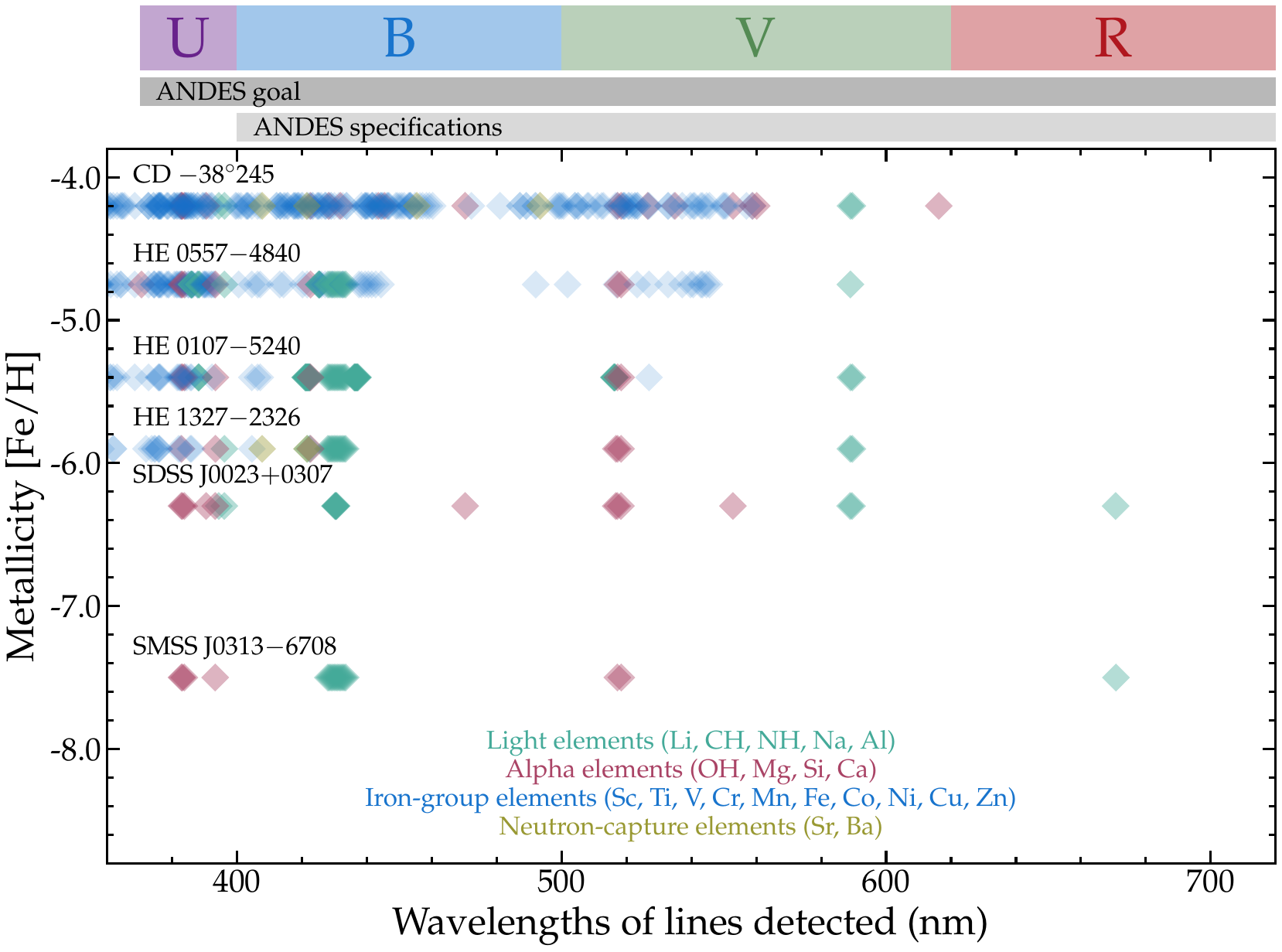}
\caption{
Wavelengths of absorption lines detected in the optical spectra of six stars known at present with [Fe/H] $< -4$. 
Each point represents one line in one star, and different elements are marked by different colors.
The design of ANDES will enable the detection of the vast majority of these lines, covering all major nucleosynthesis groups, in many more stars that retain the fossil records of metals produced by the first stars.
\textit{%
Figure modified from
\citet{roederer19baas}. 
}
}
\label{fig:firststars}
\end{figure}

\begin{figure}[h]%
\centering
\includegraphics[width=0.95\textwidth]{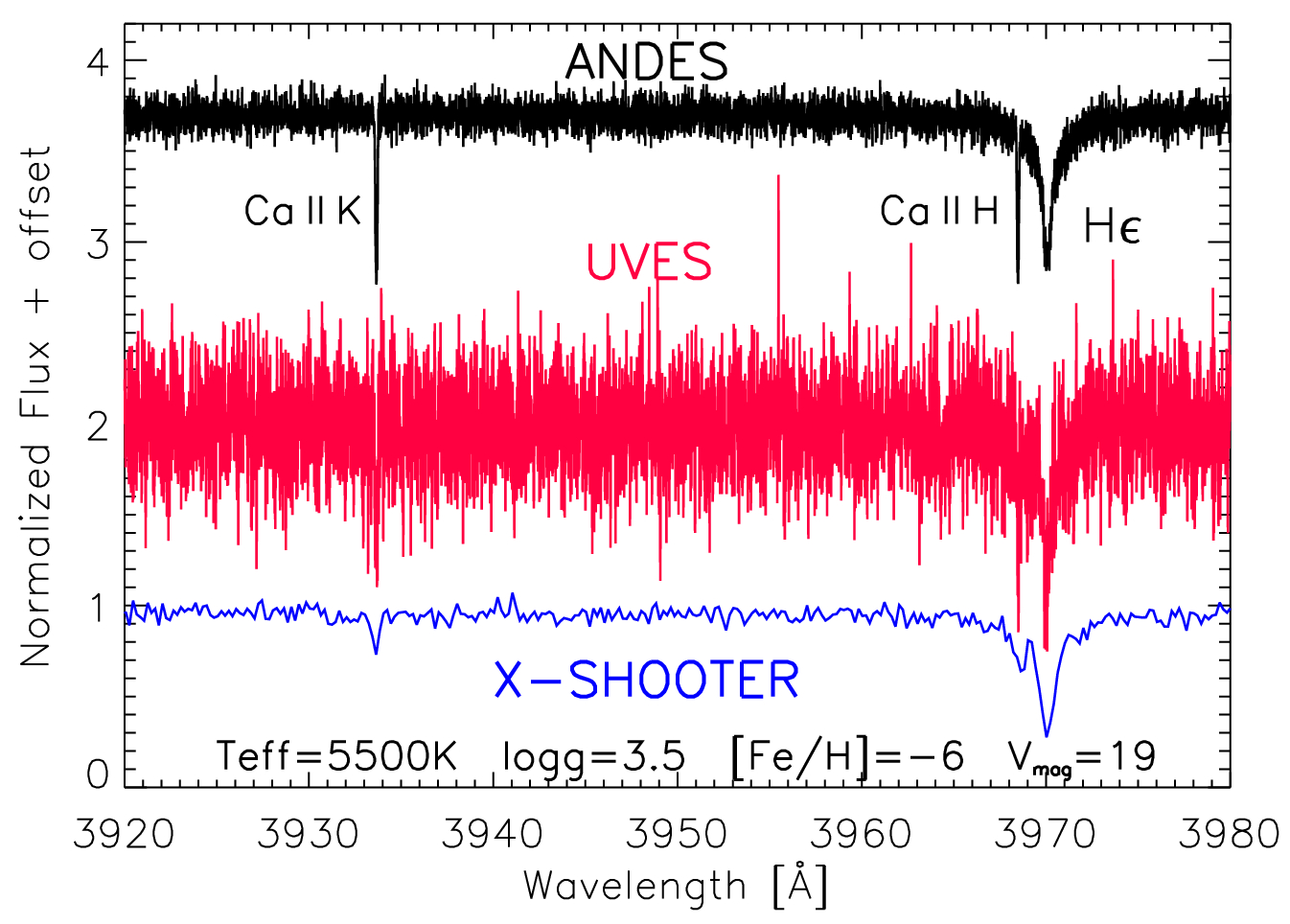}
\caption{A simulated 1-hour spectrum of a subgiant second-generation star 
with a total metallicity of $10^{-6}$ times the Solar metallicity 
in the $U$ channel around Ca~\textsc{ii} H~\&~K lines. 
Random noise has been added to simulate the S/N for ANDES (black), UVES (red), and X-SHOOTER (blue). 
}
\label{fig:zero_sim}
\end{figure}

The key observational signatures to be examined include numerous metal absorption lines, spanning 370 to 670~nm, in the most metal-poor stars.
Figure~\ref{fig:firststars} illustrates the wavelengths of these lines in known metal-poor stars, and it highlights the importance of the ANDES $U$ and $B$ channels to detect many of these lines.
About 40\% more lines would be detectable in these stars with the inclusion of the $U$ channel.
High-resolution and high-S/N spectra are necessary to enable the detection of weak metal lines or place meaningful upper limits on non-detections of these lines.

Similar work can be conducted with existing high-resolution spectrographs on 4--10~m class telescopes, but only modest-size samples of targets are within reach of high-resolution optical spectrographs on these telescopes.
Figure \ref{fig:zero_sim} shows simulated spectra around the Ca~\textsc{ii} H \& K lines for a subgiant star at a distance of $\approx 20$~kpc where the total amount of elements heavier than helium is $< 10^{-6}$ Solar.
Observations using X-SHOOTER and UVES at the VLT would lack sufficient S/N and spectral resolution to conclusively identify the unique nature of this hypothetical star, whereas ANDES at the ELT could easily resolve the stellar Ca~\textsc{ii} lines.
Ongoing and planned surveys---including ones conducted with
4MOST,
DECAM,
DESI,
Gaia,
LAMOST,
MOONS,
Pristine,
the Rubin Telescope,
SkyMapper, 
S-PLUS,
and more---%
will identify thousands of targets with $V$ magnitudes fainter than 18 that are too faint for high-resolution spectroscopic follow-up with the current generation of optical telescopes.
The large collecting area of the ELT will greatly expand the samples of first-star candidates and their descendants that can be observed in the Milky Way disk, halo, bulge, and Local Group dwarf galaxies.

\subsection{Stellar populations in the Galactic bulge}
\label{sec:bulge}

The age distribution of stars in the Galactic bulge 
is a key ingredient to understanding the processes that shaped the formation and evolution of the bulge \citep{barbuy18}.
Observational evidence has, however, presented conflicting views on this subject.
Photometric studies have found that the bulge is an almost entirely old stellar population \citep[e.g.,][]{zoccali2003,clarkson2008,renzini2018}.
In contrast, studies based on isochrone ages of dwarf, turnoff, and subgiant stars magnified by gravitational microlensing events have reported that the fraction of young- and intermediate-age stars may be as high as 50\% at super-solar metallicities \citep{bensby2013,bensby2017}.  

The large collecting area of the ELT will enable ANDES to obtain high-resolution spectra of turnoff and subgiant stars in the bulge, even when they are not being magnified in gravitational microlensing events. 
Turnoff and subgiant stars located in the bulge have apparent magnitudes of 18 $\leq V \leq$ 19.
ANDES will be able to obtain a spectrum with $R =$ 100,000
(50,000) and $S/N\approx$ 30--40 (50--60) for a bulge turnoff star in 1--2~hr.

The optical spectrum (500--900~nm) of each star contains numerous lines of Fe and $\alpha$ elements---including O, Mg, Si, Ca, and Ti---that are necessary to determine parameters, metallicities, and other detailed chemical abundances.
With this information, stellar ages can be inferred with a precision of about $25$\%, depending on the evolutionary state of the star, from theoretical isochrones \citep[][]{bensby2017}.
By systematically mapping the age distribution of the bulge across a wide range of Galactic longitude and latitude, ANDES would enable novel constraints on the the formation and nature of the bulge and its connections to other stellar populations.

ANDES will also be uniquely capable of studying the oldest stellar populations buried deep within the bulge for the first time.
In the $\Lambda$CDM paradigm, galaxies like the Milky Way formed the largest concentration of old stars in their innermost regions (e.g., \citealt{tumlinson10,Salvadori2010,elbadry18,pagnini2023}). 
The old stars that survive in these innermost regions today are inaccessible to optical or UV spectroscopy, but they can be observed in the NIR spectral range.
High-resolution and high-S/N spectroscopy covering $\sim$~950--2400~nm is generally not possible for large samples of candidate metal-poor stars within the innermost few kpc of the Galaxy with existing facilities, because the extinction is so high.
Studies using ANDES would surpass previous efforts that have helped to establish the potential of this approach, such as the APOGEE \citep{queiroz21}, the EMBLA \citep{howes15}, and the PIGS \citep{arentsen2020} surveys.
There would be synergy with the MOONS survey 
and ELT's HARMONI instrument.
Their large multiplexing capabilities could identify the most promising targets for ANDES follow-up to derive the compositions of a few dozen elements that trace the nucleosynthesis of the earliest generations of stars in the bulge.
Key spectral lines of interest include 
Na~\textsc{i} 1638.8~nm, 
Mg~\textsc{i} 1208.3~nm, 
Ti~\textsc{ii} 1587.3~nm, 
Fe~\textsc{i} 1287.9~nm, 
Fe~\textsc{i} 1507.7~nm, 
Zn~\textsc{i} 1105.4~nm, 
Sr~\textsc{ii} 1032.7~nm, 
Ce~\textsc{ii} 1578.4~nm, and
Eu~\textsc{ii} 1016.5~nm
\citep{cunha17,matsunaga20,smith21,fanelli21}.

\subsection{Reaching the edge of the Galactic disk with open clusters}
\label{sec:openclusters}

Open clusters in the Galactic disk span a wide distribution of ages and distances, and the properties of their stars can be determined both accurately and precisely.
These features place open clusters among the best tracers of the properties and stellar populations of the disk.
For example, our knowledge of the shapes of chemical abundance gradients in the outer part of the Galaxy is still unsettled, and these gradients hold important implications for understanding the formation of the Galactic disk and the impact of stellar migration \citep[e.g.,][]{magrini23}.

Studies using data from the Gaia mission have characterized the ages and distances of all known open clusters, and these data have revealed many new ones. Large spectroscopic surveys such as Gaia-ESO, APOGEE, and GALAH have derived the chemical properties of some of them, but these surveys---and future ones with WEAVE and 4MOST---are limited in magnitude.
These surveys are unable to chemically characterize clusters across the full radial extent of the Galactic disk.

\begin{figure}[h]%
\centering
\includegraphics[width=0.8\textwidth]{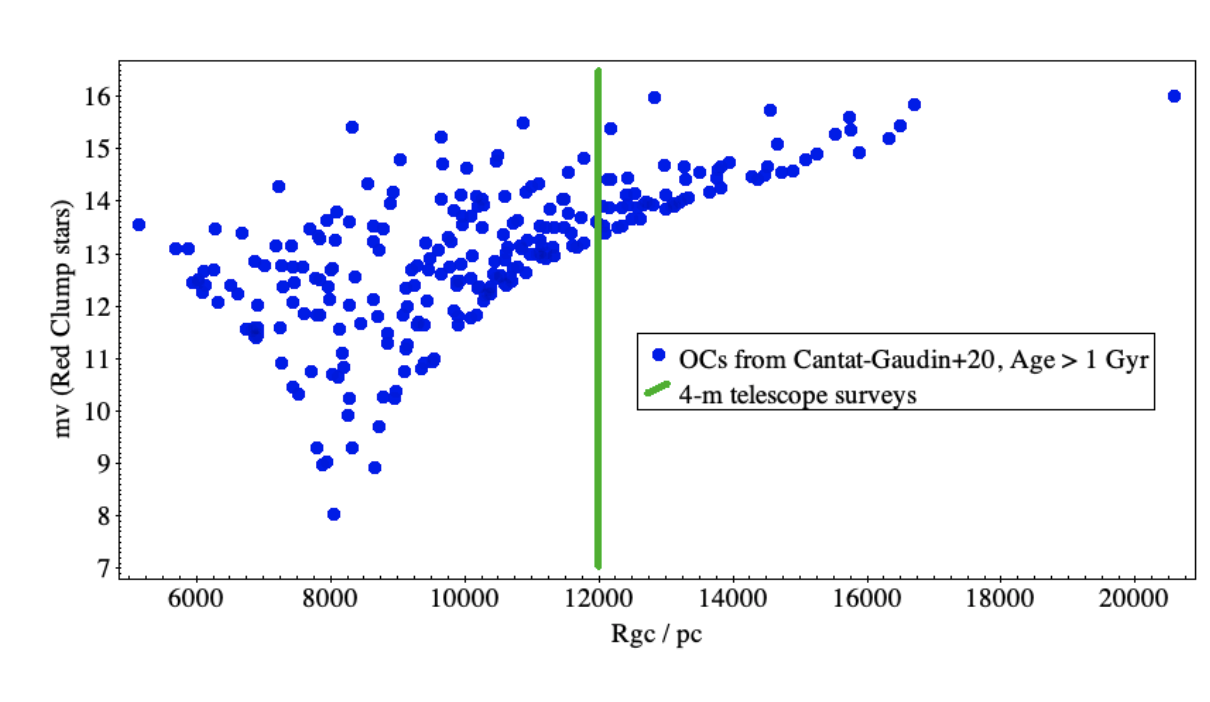}
\caption{Distribution of apparent $V$ magnitudes of red clump stars in open clusters as a function of their galactocentric radius, R$_{\rm gc}$.
Distances and ages are adopted from \citet{cantat20}.
ANDES will be able to collect high-resolution spectra to chemically characterize stars in clusters all the way to the outer edge of the Galactic disk, which remains out of reach of current surveys and large optical telescopes.
}
\label{fig:rc}
\end{figure}

ANDES will overcome these limitations and be able to collect high-resolution spectra of open cluster stars all the way to the outer edge of the Galactic disk.
In Fig.~\ref{fig:rc}, we show the distribution of the apparent $V$ magnitudes, $m_{V}$, of red clump stars in open clusters with age $>$~1~Gyr. 
ANDES can easily reach the farthest clusters with short exposure times (1800~s for S/N~$\sim$~120 at 650~nm with $R =$~100,000 for a star with $m_{V} = 16$),
such that the chemical abundances of a few dozen elements can be characterized in 4--5~stars in $<$~2~hr per cluster.

\subsection{The mysterious nature of M subdwarfs}
\label{sec:mdwarfs}

Cool subdwarf stars have lifetimes much longer than the Hubble time \citep{Laughlin1997}.
These faint, low-mass stars are relatively rare in the solar neighborhood, but they are predicted to be the most numerous stellar community within the Milky Way’s halo \citep{Bochanski2013}.
Studies of cool subdwarf spectra show that they are metal poor when compared with the more common cool dwarfs with near-solar metallicity \citep{Gizis1997, Lepine2007}. 
The nature of subdwarfs has been explored over the last six decades by a large number of authors \citep{Sandage1959, Hartwick1984, Gizis1997, Zhang2019},
but fundamental questions about their origins and kinematics remain.
For example, searches for cool subdwarfs have been limited to high Galactic latitudes, which favors the detection of halo and thick disk objects.
Whether or not a cool subdwarf population exists within the Galactic thin disk remains an open question for observations and a challenge for stellar evolution theory.

ANDES will be capable of collecting spectra of cool subdwarfs in unprecedented detail.
The broad spectral coverage of ANDES 
will enable 
derivation of accurate atmospheric parameters and chemical compositions of cool subdwarfs in all Galactic populations where they are found.
These quantities are essential ingredients to understand the origin and nature of this mysterious class of objects.

\subsection{Resolved stellar populations across the dwarf galaxies in the Local Group}
\label{sec:dwarfgalaxies}

The Local Universe provides a unique window into the process of hierarchical mass assembly on all scales.
The Local Group of galaxies spans a range of more than a million in stellar mass, from the Large Magellanic Cloud stellar mass of $\sim3 \times 10^{9}$~\msun\ to numerous dwarf galaxies with stellar masses $< 10^{3}$~\msun\ (e.g., \citealt{McConnachie2012, Simon2019}), and it
includes many gas-poor, isolated systems and several gas-rich interacting galaxies.
These galaxies will be prime targets of the next generation of multi-object spectrographs, such as 4MOST 
and MOONS.~ 
Those surveys will identify thousands of stars in these galaxies that are too faint for high-resolution and high-S/N optical spectroscopy with existing facilities.
Yet those stars will hold the promise of unlocking our understanding of galaxy formation and evolution
at the low mass scale 
\citep{Wheeler2015, Applebaum2021}
and the epochs of the earliest chemical enrichment
\citep{frebel2015}.


Most Local Group dwarf galaxies are intrinsically metal-poor with a dominant old stellar population, so they are ideal fossils with which to study the imprints of the first stars and discover the nature and properties of the earliest nucleosynthesis events \citep{Ishigaki2018, Kobayashi2020, skuladottir21}.  
The fundamental physical conditions of major nucleosynthetic sites remain obscure, even after decades of study. 
The key to breaking degeneracies in this field is to investigate the chemical compositions of stars in galaxies with a variety of star formation and chemical enrichment histories.
For example, these chemical compositions provide insight into the progenitors of Type~Ia supernovae \citep{Maoz2014},
the production of heavy elements in AGB stars 
(via the s- and i-processes; \citealt{Ritter2018, Skuladottir2020}),
the products of the kilonovae that accompany neutron star mergers (via the r-process; e.g., \citealt{Kasen2017,ji23ret2}), and 
the source(s) of the extreme carbon enhancement observed in many extremely metal-poor stars \citep{norris13cemp, Salvadori2015,Jeon2021}.

Furthermore, some ultra faint dwarf galaxies are thought to be surviving first galaxies \citep{BovillRicotti2009, SalvadoriFerrara2009, Frebel2012, Starkenburg2017}. 
They offer a complementary perspective on high redshift systems, such as the ones expected to be observed with JWST.~
Theoretical simulations 
based on $\Lambda$CDM theory predict a large number of low-mass satellites for galaxies with masses comparable to the Magellanic Clouds.
Unfortunately, most of these satellites are too faint to be directly detected, and others will have already dissolved to unrecognizable levels after merger events with more massive dwarfs.
Fortunately, stars with peculiar abundances in the more massive dwarfs, such as the Magellanic Clouds and the Sagittarius dwarf, retain chemical signatures of their now-dissolved birth galaxies
\citep{Chiti2020,Reggiani21,Mucciarelli2023}.
The unprecedented statistical scales of stars observed by 4MOST and MOONS around the Magellanic Clouds, for example, will reveal numerous rare stars in these systems with [Fe/H]$<-2.5$.





\begin{figure}[h]%
\centering
\includegraphics[width=0.55\textwidth]{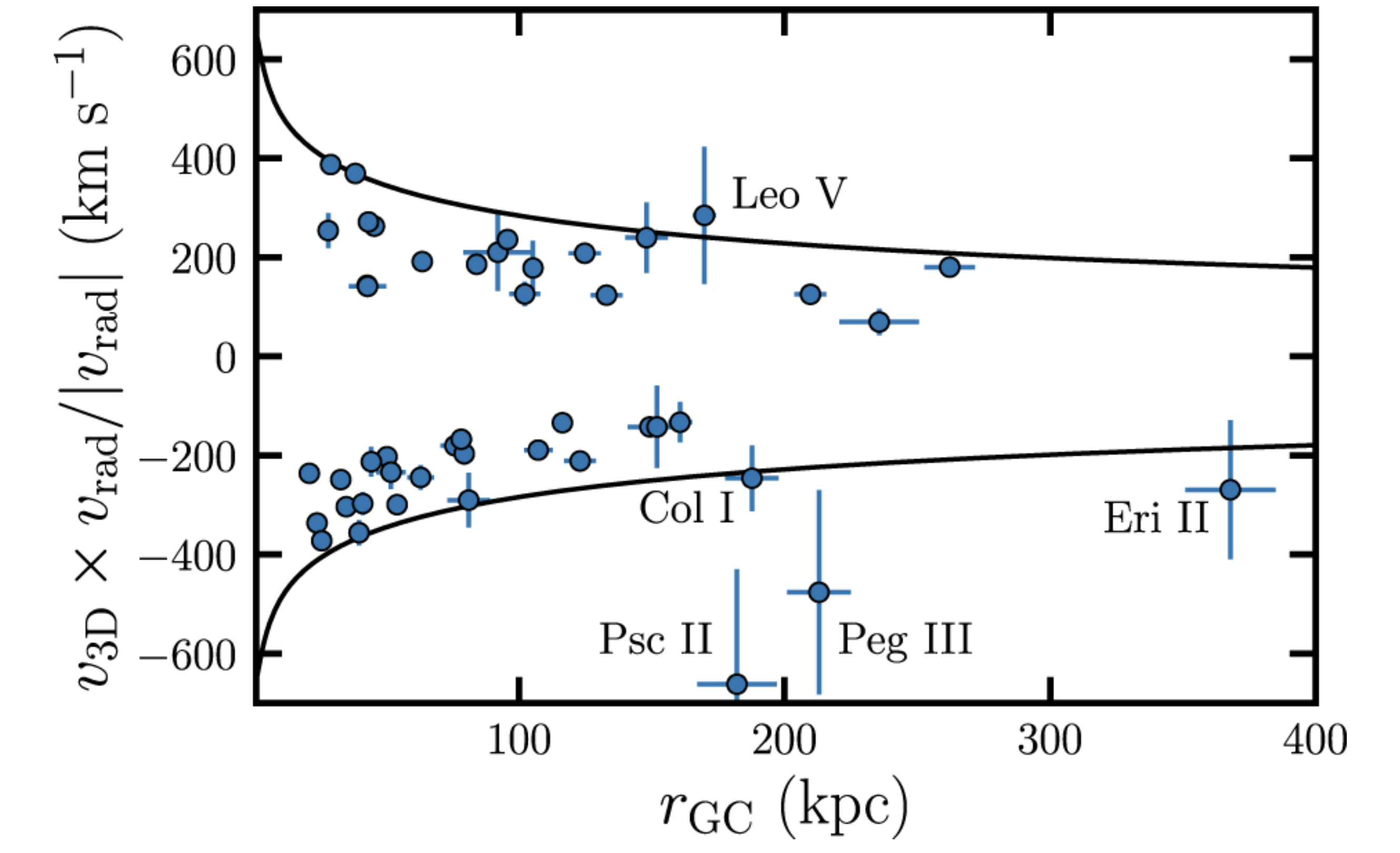}
\hfill
\includegraphics[width=0.4\textwidth]{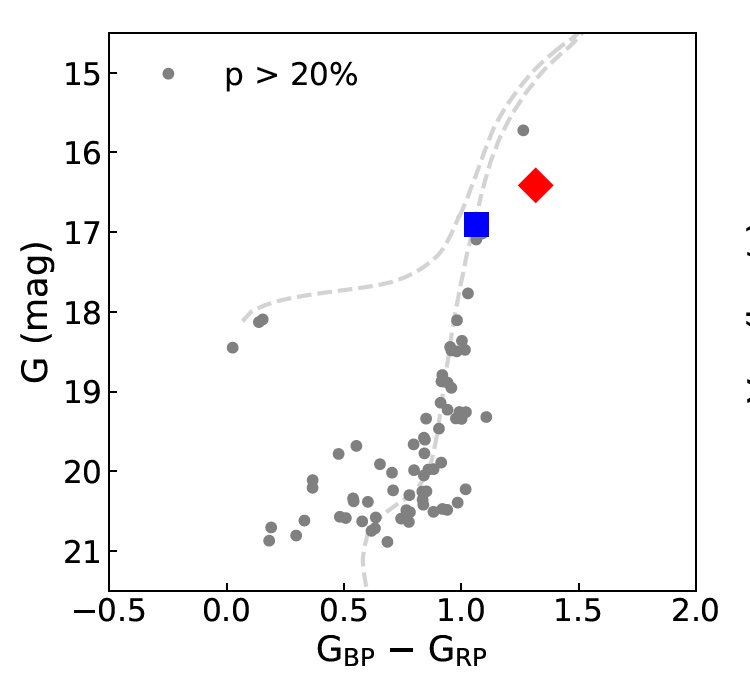}
\caption{Left:\ phase-space diagram for the Milky Way dwarf galaxies (from \citealt{Pace2022}), where the x-axis shows the distance, $r$, to the Galactic center (GC) and the y-axis shows the 3-dimensional velocity relative to the GC.~
Positive (negative) velocities indicate that the dSph is moving away from (toward) the GC, and 
the black lines represent the escape velocity of the MW.~
Right:\ Gaia color-magnitude diagram for the ultra-faint dwarf galaxy \mbox{Reticulum~II}, showing stars with $>20\%$ membership probability, $p$ \citep{Jensen2023}.
The blue square and red diamond show two bright stars recently observed using the GHOST spectrograph on the 
8~m Gemini-South Telescope \citep{Hayes2023}.
Most stars in \mbox{Reticulum~II} are fainter than $G$ = 19,
which is one of the nearest dwarf galaxies,
demonstrating the need for ANDES to collect 
high-resolution spectra of most stars in Local Group dwarf galaxies.
}
\label{fig:mwdwarfs}
\end{figure}

ANDES will provide the opportunity to perform high-resolution spectroscopic follow-up of all of these interesting stars that will be discovered in the Milky Way satellites in the next few decades.
Virtually all stars in Local Group galaxies beyond the Milky Way have Gaia magnitudes $G \gtrsim19$ 
(Figure~\ref{fig:mwdwarfs}),
so high-resolution spectroscopy of their stars will require the collecting area of telescopes on the scale of the ELT.
Many of the spectral lines necessary for these analyses are 
found at relatively blue wavelengths ($\lesssim$~460~nm),
where detector efficiencies and spectrograph throughputs decrease.
Even relatively bright stars in dwarf galaxies will require the higher throughput at all wavelengths that will be available with ANDES.

\subsection{Evolved stars in other galaxies}
\label{sec:evolvedstars}

Evolved cool stars of various masses are major cosmic engines, providing strong mechanical and radiative feedback on their host environment \citep{langer2012araa} through stellar winds and supernova mass ejections \citep{2018A&ARv..26....1H}. 
These M-type stars of are among the largest and most luminous stars in the Universe. The brightest M supergiants should be excellent extragalactic distance indicators for a wide range of galaxy types and luminosities \citep[e.g.,][]{1983ApJ...269..335H} and they provide robust chemical abundance estimates that constrain models of galaxy formation and chemical evolution within the Milky Way \citep[e.g., with JWST;][]{2018ApJ...867..155L} and nearby galaxies \citep{2017ApJ...847..112D}.
Massive evolved stars have been used to examine the metal content and distribution in Local Universe galaxies out to distances of $\sim 20$~Mpc using $J$-band spectroscopy \citep{2015ApJ...805..182G,2017MNRAS.468..492P,2022ApJ...940...32B} and to study the distances and properties of Galactic, young massive star clusters 
\citep{2019MNRAS.486L..10D,2022A&A...661L...1C,fanelli22a,fanelli22b,2022MNRAS.516.1289N}.
Metallicities of individual massive stars, asymptotic giant branch stars, and red supergiant stars that are bright in the mid infrared can be estimated at distances of tens of Mpc \citep{2011A&A...527A..50E}. 

\begin{figure}
  \centering
    \begin{tabular}{cc}
  \includegraphics[width=0.415\textwidth]{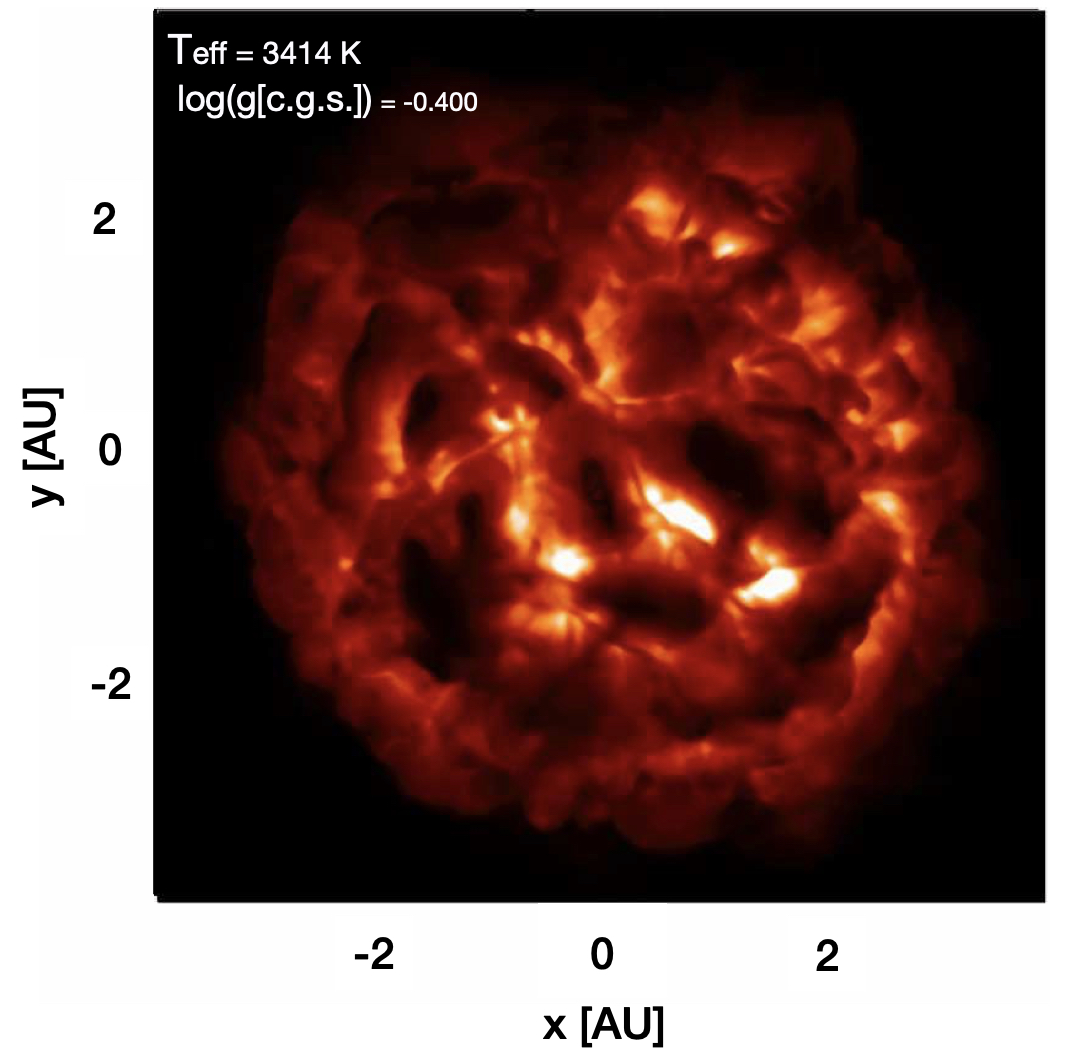}
   \includegraphics[width=0.54\hsize]{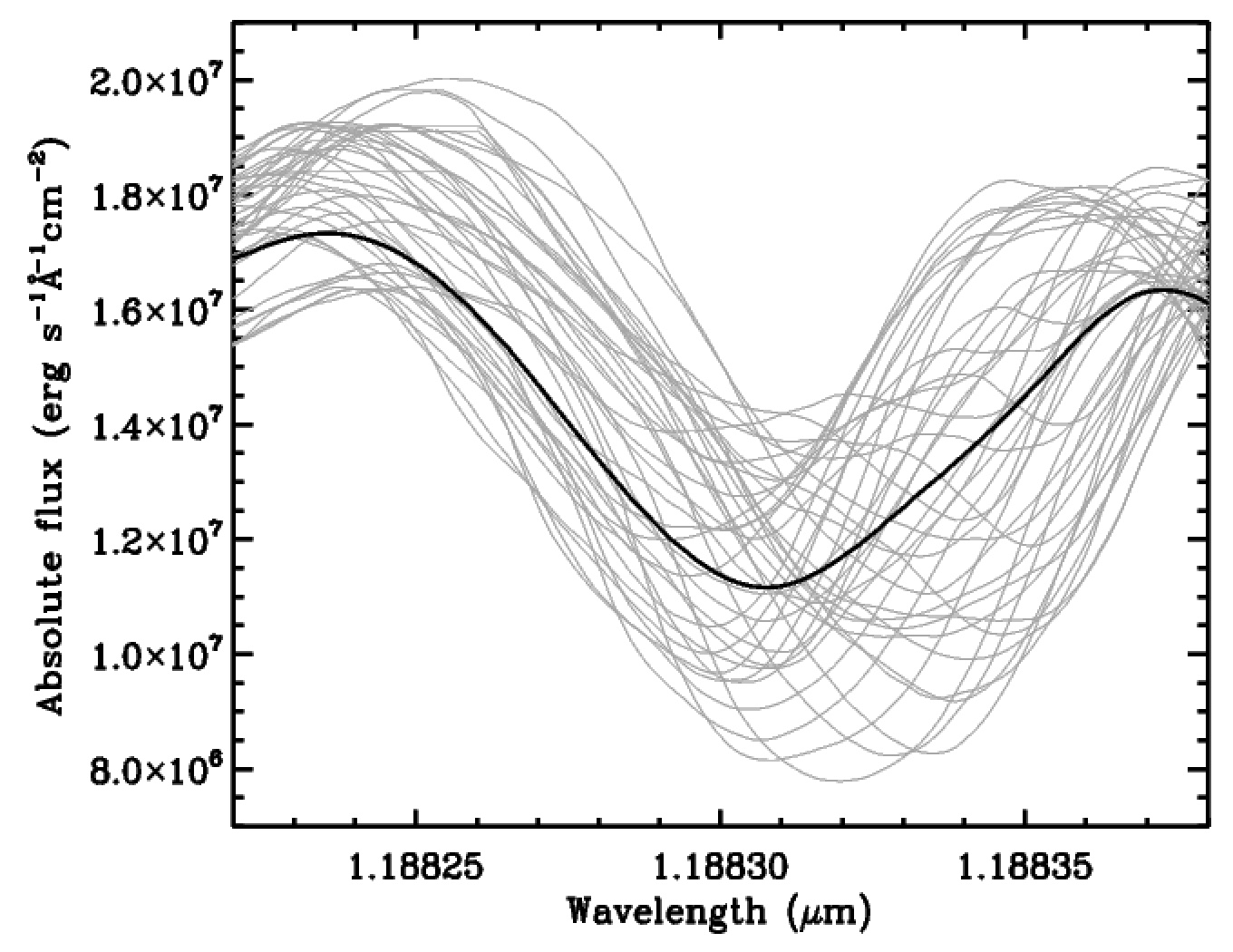}
\end{tabular}
\caption{\emph{Left:} Example of synthetic intensity map from a 3D radiation-hydrodynamics simulation of stellar convection 
\citep{2022A&A...661L...1C}. 
\emph{Right:} zoom-in view around an Fe~\textsc{i} line in the J-band taken from the same simulation, at the spectral resolving power of ANDES.
The light gray displays spectra at timesteps offset by $\sim$~21~d, while the solid black line is the temporal average.
ANDES will be capable of resolving these differences in the line profiles.
}
\label{fig:RSG}       
\end{figure}

However, these stars are a cauldron of complexity for all the physics involved from the mixing in the interior towards the transport into the highly dynamical surfaces (Fig.~\ref{fig:RSG}, left), and, eventually the extremely vigorous (still not completely understood) mass loss mechanism \citep{2018A&ARv..26....1H}. 
The emerging stellar spectra 
and intensity map are characterised by broad, asymmetric and shifted lines (Fig.~\ref{fig:RSG}, right) with deviations steamed by a temporal dependent and episodic convection pattern consisting of large granules and (super)-sonic velocities \citep[e.g., ][]{2008AJ....135.1450G}. Theoretically confirmed by state-of-the-art numerical simulations \citep[e.g., ][]{2011A&A...535A..22C}.



ANDES offers a new opportunity to obtain high-resolution spectroscopy of evolved stars beyond the Local Group.
Spectral diagnostics include atomic lines of Mg, Si, Ti, and Fe in the J-band, where we can obtain uncertainties on the abundances of $\pm$0.1 dex (at solar metallicities) with S/N $> 55$ at resolving power lower than 10000 \citep{2011A&A...527A..50E}. However, due to the complex dynamics of those stars, higher spectral resolution are needed to resolve the shape and temporal shifts of the lines (Fig.~\ref{fig:RSG}, right panel). This represents an important step to address fundamental questions about the evolution of massive stars, their environments, as well as their metallicity in other galaxies \citep{2011BSRSL..80..456E}.

\subsection{Calibrating properties of Cepheid variables for the cosmic distance ladder}
\label{sec:cepheids}

Cepheid variables are the backbone of the extragalactic distance ladder. 
The discovery of the accelerating expansion of the Universe, which was awarded the Nobel Prize in 2011, is based on our understanding of Cepheid variables \citep{riess22}.
However, there is currently a 5-$\sigma$ tension between the expansion rate of the universe, $H_0$, depending on whether it is derived from the cosmic microwave background or from the distance ladder \citep{divalentino21}.
If confirmed, this tension would indicate that the $\Lambda$CDM model of the universe should be refined.

ANDES observations will enable a new route to determine $H_0$ using the Baade-Wesselink method of Cepheid distance determination \citep{kervella04a,nardetto23}.
This method measures distances to Cepheid variables as follows. 
The variation of the stellar angular diameter, which can be obtained from surface brightness-color relations or by interferometric measurements, is compared to the variation of the linear diameter, which can be derived from radial velocity measurements multiplied by the projection factor. The distance to the Cepheid can then be obtained by dividing the linear and angular diameters. This method has the potential to test the tension in $H_{0}$ by determining the distances to individual Cepheids in the Local Group and beyond, from which the period-luminosity-metallicity relations in various galaxies can be calibrated consistently.

This method takes advantage of the unmatched capabilities of ANDES in three ways.
First, ANDES will enable the study of the dynamical structure of the atmospheres of extragalactic Cepheids \citep[e.g.,][]{2006A&A...453..309N,2017A&A...597A..73N}.
Secondly, ANDES will introduce new anchors as constraints on $H_{0}$ by providing Baade-Wesselink distances \citep{2011A&A...534A..94S,2011A&A...534A..95S,2017A&A...608A..18G,2021A&A...656A.102T} to individual Cepheids throughout the Local Group, including ones in the Large and Small Magellanic Clouds, IC~1613, and NGC~6822.
Finally, ANDES will enable the derivation of Cepheid metallicities and other chemical abundances, such as $\alpha$-element abundances, which will help to calibrate and interpret the dependence of the period-luminosity relationship on chemical composition \citep{2005A&A...429L..37R,2008A&A...488..731R}.

\section{The Star-Planet Connection}
\label{sec:starplanet}

\subsection{Stellar chemical composition as a driver of exoplanet formation}
\label{sec:hoststarchemistry}

The field of exoplanets has been rapidly growing since the discovery of the first exoplanet around a main-sequence star almost 30 years ago \citep{Mayor95}.
As the sample of exoplanets has grown, it has been possible to statistically compare their properties with those of their host stars---including age, mass, chemical composition, effective temperature ($T_{\rm eff}$), surface gravity ($\log~g$), and more.
High-resolution spectroscopy with ANDES will advance our understanding of these connections.

A higher occurrence frequency of Jupiter-like planets has been observed around high-metallicity stars than low-metallicity stars, supporting the scenario that giant planets formed from the core accretion process \citep[e.g.,][]{Adibekyan19, Osborn20}. 
Rocky Earth-like planets, in contrast, exhibit a weaker correlation with host star metallicity \citep{Mordasini12, Kutra21}. 
The abundances of many elements, in addition to iron, are predicted to affect planet formation, 
so detailed chemical abundance studies of the stellar hosts are required to formulate a more complete picture of how planets form 
\citep[e.g.,][]{Adibekyan12, Hinkel14, DelgadoMena17, Biazzo22}.
The abundance of elements such as carbon, nitrogen, oxygen, and sulfur in exoplanet hosts also provides fundamental constraints on the mechanisms of planetary formation and migration in protoplanetary disks.
ANDES will play a fundamental role in characterizing the chemical composition of exoplanet hosts.
High-resolution spectroscopy across a broad spectral range (0.35--2.45 $\mu$m) of many stars will reveal how the abundances of individual elements and isotopes map to planetary properties.

The atmospheres of host stars are covered with complex, stochastic patterns associated with convective heat transport, such as granulation, that can severely complicate the interpretation of exoplanet spectra \citep{2017A&A...597A..94C,2019A&A...631A.100C}. The impact of granulation on planetary signal extraction depends on the stellar parameters and stellar metallicity.
The wide wavelength coverage of ANDES will be well suited to disentangle these effects. 
Molecules such as CO, H$_2$O, and HCN are readily observable in the NIR spectra of main-sequence and M-dwarf stars, for example, and TiO, VO, and numerous atomic lines are found in the optical spectra of these stars.
All of these lines provide diagnostic abilities critical to characterizing the internal stellar dynamics and extracting reliable results from planetary atmospheres.

In addition to the chemical composition affecting the architecture of planetary systems, the formation of planets will have an impact on the composition of the host star.
Formation of the rocky planets in the Solar System may be responsible for the subtle, but measurable, differences between the composition of the Sun and most nearby solar twins \citep{2009ApJ...704L..66M}, which exhibit clear correlations with condensation temperature.
The opposite behavior may occur when rocky planets, after orbit destabilization due to dynamical interactions with other bodies, are accreted by the host star \citep{2021NatAs...5.1163S}.

Chemical abundances can be derived from atomic lines and molecular bands, including the [O~\textsc{i}] line at 630.0~nm, the [C~\textsc{i}] line at 872.7~nm, C~\textsc{i} transitions at 505.2~nm and 538.0~nm, and the S~\textsc{i} line at 675.7~nm.
Planetary host stars are generally close and bright, and S/N ratios of 100--150 \AA$^{-1}$ yield excellent chemical characterisations. 
The large collecting area of the ELT will enable ANDES to greatly expand the number of exoplanet host stars where studies such as these can be conducted.
ANDES spectra will complement data collected by the Ariel satellite \citep{tinetti22}, which will characterize the atmospheres of exoplanets and their stellar hosts at low spectral resolution
\citep{Danielski22, Magrini22}.

\subsection{Star-planet interactions}
\label{sec:spi}

Many known exoplanets orbit quite close to their host stars, with orbital periods of only a few days. This close proximity can give rise to star-planet interactions (SPI), tidal or magnetic interactions through which the exoplanet influences its host star.

\textit{Tidal SPI} refers to the tidal influence a planet has on its host star, which can lead to an exchange of angular momentum between the planet's orbit and the stellar spin. 
Manifestations of tidal SPI include elevated stellar rotation \citep{Pont2009, Brown2011}, enhanced overall magnetic activity \citep{Kashyap2008, Poppenhaeger2014, Ilic2022}, and disruption of magnetic activity
\citep{Pillitteri2014}. 

\textit{Magnetic SPI} concerns the interplay between the stellar magnetic field and a close-in orbiting planet, with or without a magnetic field \citep{Strugarek2014}. 
Magnetic SPI can yield detectable radio emission via the electron-cyclotron maser 
instability. 
In the optical spectral range, the magnetic interaction is thought to enhance local activity in the star, but
these signals have not been explored as well at optical wavelengths as they have in the radio. 
The SPI indicators would be modulated in both wavelength domains with the planet’s orbit, not the stellar rotation period.
Early indications of magnetic SPI in the optical remain unconfirmed, suggesting a mechanism that does not operate continuously \citep{Shkolnik2005,Shkolnik2008}.
Several recent studies have revitalized the field by detecting radio emission from nearby stars (Proxima~Cen, \citealt{PerezTorres2021}; YZ~Cet, \citealt{Pineda2023}) at certain phases of the orbit of their planets, making these systems the best candidates to be nominated as the first detection of SPI at radio wavelengths,
and obvious candidates for optical followup with ANDES.

Observations constraining both tidal SPI and magnetic SPI 
are critical to our understanding of exoplanet habitability.
For example, ANDES observations of stellar spectral signatures could reveal the rotation periods of exoplanets and their magnetic field strengths,
which are thought to be an important ingredient in exoplanet habitability \citep{2014A&A...570A..99S, 2017ApJ...844L..13G, 2017ApJ...837L..26D}.
The high spectral resolution and broad wavelength coverage
of ANDES will probe stellar rotation through line broadening, with a wealth of stellar absorption lines particularly in the UBV and RIZ channels; it will measure chromospheric activity through simultaneous observations of the Ca~\textsc{ii} H~\&~K lines, the H$\alpha$ line, the Ca~\textsc{ii} NIR triplet, and the He~\textsc{i} NIR triplet; and characterize properties of the upper convection zone of stars through resolved line profiles.
The unmatched light-collecting area of the ELT and the broad spectral coverage provided by ANDES will enable---for the first time---searches for the rotation period of exoplanets based on different methods, including rotational broadening and Doppler shift of lines in the planetary atmospheres \citep{Snellen2014}
and analysis of starlight reflected from the planet
\citep{Li2022, Aizawa2020}. 

ANDES will enable greater sensitivity of these effects in much larger samples of exoplanet host stars than is accessible at present.
The combination of the planet rotation period, obtained from ANDES observations, and the stellar rotation and planet orbital periodicities, obtained from all-sky photometric space missions, such as TESS and PLATO, will produce a unique portrait of SPI.~

\section{Summary}
\label{sec:summary}

ANDES will enable a wide range of potentially transformative science across the field of stellar astrophysics.
These exciting opportunities include expanding our understanding of the physics of how stars work, the unique roles that stars play in shaping the formation and evolution of galaxies, and the intimate connections between stars and their systems of planets.
The science cases presented here often rely on more than one of the wide diversity of the proposed capabilities of ANDES, and they represent only a fraction of the potential science that ANDES could enable while in operation on the ELT in the decades ahead.
The takeaway message may be summarized as follows:\ 
\textit{Let's build ANDES together, and the potential for transformative stellar astrophysics science will be virtually unlimited.}

\bmhead{Acknowledgments}


IUR acknowledges support from United States National Science Foundation grants AST~1815403 and AST~2205847.
CAP is thankful to the Spanish Ministry of Science and Innovation (MICINN) for funding under the project PID2020-117493GB-I00.
TB acknowledges support from project grant No.\ 2018-04857 from the Swedish Research Council.
E.M.A.G.\ acknowledges support from the German \textit{Leibniz-Gemeinschaft} under project number P67/2018.
RM acknowledges support from the Science and Technology Facilities Council
(STFC), by the ERC Advanced Grant 695671 ‘QUENCH,’ and by the UKRI Frontier Research grant RISEandFALL. 
RM also acknowledges funding from a research professorship from the Royal Society.
AC acknowledges support from the French National Research Agency (ANR) funded project PEPPER (ANR-20-CE31-0002).
AJK acknowledges support by the Swedish National Space Agency (SNSA).~
VA is supported by FCT (Funda\c{c}\~ao para a Ci\^encia e Tecnologia) through national funds by the following grants: UIDB/04434/2020, UIDP/04434/2020, and 2022.06962.PTDC.~
CRL and PJA acknowledge financial support from the Agencia Estatal de Investigaci\'on of the Ministerio de Ciencia e Innovaci\'on through projects PID2019-109522GB-C52, PID2022-137241NB-C43 and the Severo Ochoa grant CEX2021-001131-S all funded by MCIN/AEI/10.13039/501100011033.
DR acknowledges support from the Italian National Institute for Astrophysics through Theory Grant 2022, Fu.~Ob.~1.05.12.06.08.
JRM acknowledges continuous support from the Universidade Federal do Rio Grande do Norte and Brazilian Agencies CNPq and CAPES.

\textbf{Author contributions.}

ADD HERE FROM THE 
\href{https://docs.google.com/document/d/1XKDO1oTQQX92dn5-QkLPs1xZdBwAKEqWf-V6ygcKl_0/edit}{GOOGLE DOC}

\bibliography{wg2}

\end{document}